\documentclass[twocolumn]{aastex631}
\usepackage{xcolor}

\usepackage{natbib}
\usepackage{amsmath}
\usepackage{enumitem}
\usepackage{listings}
\usepackage{appendix}
\usepackage{float}

\begin{document}

\title{Predictions of the LSST Solar System Yield: Near-Earth Objects, Main Belt Asteroids, Jupiter Trojans, and Trans-Neptunian Objects}
\author[0009-0005-5452-0671]{Jacob A. Kurlander} 
\correspondingauthor{Jacob A. Kurlander}
\email{jkurla@uw.edu}
\affiliation{DiRAC Institute and the Department of Astronomy, University of Washington, 3910 15th Ave NE, Seattle, WA 98195, USA}
\author[0000-0003-0743-9422]{Pedro H. Bernardinelli}
\altaffiliation{DiRAC Postdoctoral Fellow}
\affiliation{DiRAC Institute and the Department of Astronomy, University of Washington, 3910 15th Ave NE, Seattle, WA 98195, USA}

\author[0000-0003-4365-1455]{Megan E. Schwamb}
\affiliation{Astrophysics Research Centre, School of Mathematics and Physics, Queen's University Belfast, Belfast BT7 1NN, UK}

\author[0000-0003-1996-9252]{Mario Juri\'c}
\affiliation{DiRAC Institute and the Department of Astronomy, University of Washington, 3910 15th Ave NE, Seattle, WA 98195, USA}

\author[0000-0001-9505-1131]{Joseph Murtagh}
\affiliation{Astrophysics Research Centre, School of Mathematics and Physics, Queen's University Belfast, Belfast BT7 1NN, UK}

\author[0000-0001-7335-1715]{Colin Orion Chandler}
\affiliation{DiRAC Institute and the Department of Astronomy, University of Washington, 3910 15th Ave NE, Seattle, WA 98195, USA}
\affiliation{LSST Interdisciplinary Network for Collaboration and Computing Frameworks, 933 N. Cherry Avenue, Tucson AZ 85721}

\author[0000-0001-5930-2829]{Stephanie R. Merritt}
\affiliation{Astrophysics Research Centre, School of Mathematics and Physics, Queen's University Belfast, Belfast BT7 1NN, UK}

\author[0000-0002-4547-4301]{David Nesvorn\'y}
\affiliation{Department of Space Studies, Southwest Research Institute, 1301 Walnut St., Suite
400, Boulder, CO 80302, USA}

\author[0000-0002-6034-5452]{David Vokrouhlick\'y}
\affiliation{Astronomical Institute, Charles University, V Hole{\v{s}}ovi{\v{c}}k{\'a}ch 2, CZ-180 00 Prague 8, Czech Republic}

\author[0000-0001-5916-0031]{R. Lynne Jones}
\affiliation{Rubin Observatory, 950 N. Cherry Ave., Tucson, AZ 85719, USA}
\affiliation{Aerotek, Suite 150, 4321 Still Creek Drive, Burnaby, BC V5C6S, Canada}

\author[0000-0002-8418-4809]{Grigori Fedorets}
\affiliation{Finnish Centre for Astronomy with ESO, University of Turku, FI-20014 Turku, Finland}
\affiliation{Department of Physics, University of Helsinki, P.O. Box 64, 00014
Helsinki, Finland}

\author[0000-0002-0672-5104]{Samuel Cornwall} 
\affiliation{Dept. of Aerospace Engineering, University of Illinois at Urbana-Champaign, Urbana, IL 61801, USA}

\author[0000-0002-1139-4880]{Matthew~J.~Holman}
\affil{Center for Astrophysics | Harvard \&Smithsonian, 60 Garden St., MS 51, Cambridge, MA 02138, USA}

\author[0000-0002-1398-6302]{Siegfried Eggl}
\affiliation{Dept. of Aerospace Engineering, Grainger College of Engineering,
University of Illinois at Urbana-Champaign,
Urbana, IL 61801, USA}
\affiliation{Department of Astronomy,
University of Illinois at Urbana-Champaign,
Urbana, IL 61801, USA}
\affiliation{National Center for Supercomputing Applications,
University of Illinois at Urbana-Champaign,
Urbana, IL 61801, USA}

\author[0000-0001-6984-8411]{Drew Oldag}
\affiliation{DiRAC Institute and the Department of Astronomy, University of Washington, 3910 15th Ave NE, Seattle, WA 98195, USA}
\affiliation{LSST Interdisciplinary Network for Collaboration and Computing Frameworks, 933 N. Cherry Avenue, Tucson AZ 85721}

\author[0009-0003-3171-3118]{Maxine West}
\affiliation{DiRAC Institute and the Department of Astronomy, University of Washington, 3910 15th Ave NE, Seattle, WA 98195, USA}
\affiliation{LSST Interdisciplinary Network for Collaboration and Computing Frameworks, 933 N. Cherry Avenue, Tucson AZ 85721}

\author[0009-0009-2281-7031]{Jeremy Kubica}
\affiliation{McWilliams Center for Cosmology, Department of Physics, Carnegie Mellon University, Pittsburgh, PA 15213, USA}
\affiliation{LSST Interdisciplinary Network for Collaboration and Computing Frameworks, 933 N. Cherry Avenue, Tucson AZ 85721}

\author[0000-0003-2874-6464]{Peter Yoachim}
\affiliation{DiRAC Institute and the Department of Astronomy, University of Washington, 3910 15th Ave NE, Seattle, WA 98195, USA}

\author[0000-0001-5820-3925]{Joachim Moeyens}
\affiliation{Asteroid Institute, 20 Sunnyside Ave., Suite 427, Mill Valley, CA 94941, USA}
\affiliation{DiRAC Institute and the Department of Astronomy, University of Washington, 3910 15th Ave NE, Seattle, WA 98195, USA}

\author[0009-0006-2320-0306]{Kathleen Kiker}
\affiliation{Asteroid Institute, 20 Sunnyside Ave., Suite 427, Mill Valley, CA 94941, USA}

\author[0000-0002-8032-4528]{Laura E. Buchanan}
\affiliation{Department of Physics and Astronomy, University of Victoria, Elliott Building, 3800 Finnerty Road, Victoria, BC V8P 5C2, Canada}

\begin{abstract}
The NSF-DOE Vera C. Rubin Observatory is a new 8m-class survey facility presently being commissioned in Chile, expected to begin the 10yr-long Legacy Survey of Space and Time (LSST) by the end of 2025. Using the purpose-built \texttt{Sorcha} survey simulator \citep{Sorcha}, and near-final observing cadence, we perform the first high-fidelity simulation of LSST's solar system catalog for key small body populations. We show that the final LSST catalog will deliver over 1.1 billion observations of small bodies and raise the number of known objects to 1.27E5 near-Earth objects, 5.09E6 main belt asteroids, 1.09E5 Jupiter Trojans, and 3.70E4 trans-Neptunian objects. These represent 4-9x more objects than are presently known in each class, making LSST the largest source of data for small body science in this and the following decade. We characterize the measurements available for these populations, including orbits, \textit{griz} colors, and lightcurves, and point out science opportunities they open. Importantly, we show that $\sim 70\%$ of the main asteroid belt and more distant populations will be discovered in the first two years of the survey, making high-impact solar system science possible from very early on. We make our simulated LSST catalog publicly available, allowing researchers to test their methods on an up-to-date, representative, full-scale simulation of LSST data.

\end{abstract}

\section{Introduction}
Minor planets are of great significance to both scientific understanding of the solar system as well as planetary defense. Formed as byproducts of the processes that generated the major planets and perturbed over the last 4.5 billion years, their orbital and physical distribution now encodes the history of our Solar System \citep{ferraz1994kirkwood, michel2015asteroids}. For example, the multi-modal orbital distribution of trans-Neptunian objects (those with semi-major axis $a > 30.1$~au; TNOs) provides evidence for and constraints on the migration of Neptune \citep[e.g.][]{malhotra1993origin, ida2000orbital, tsiganis2005origin, NeptuneMigration}. Some features of the main belt keep a record of Jupiter's migration \citep[e.g.][]{MainBeltMigration, morbidelli2010evidence, walsh2012populating, 2015aste.book..493M}. And the main belt asteroids' (MBAs) color distribution as a function of orbit encodes the chemical distribution of the primordial proto-planetary disk \citep[e.g.][]{CompositionalStructure, demeo2014solar}. From a planetary defense perspective, some near-Earth objects (those with perihelion distance $q < 1.3$~au; NEOs) pose a hazard to Earth, if they have a non-negligible probability of colliding with our planet \citep{alvarez1980extraterrestrial, perna2013near, popova2013chelyabinsk}. Mitigation strategies, such as those demonstrated in the Double Asteroid Redirect Test \citep{cheng2012dart, dart_defense}, are more likely to succeed the earlier potential impactors are discovered and recognized.

As of December 2024, the Minor Planet Center (MPC) lists 1.4 million known solar system objects, the vast majority of which are members of four populations: the NEOs, MBAs, Jupiter Trojans (which orbit Jupiter's L4 and L5 points), and TNOs. Two ongoing wide-field surveys, Pan-STARRS \citep[][]{Chambers.2019} and the Catalina Sky Survey \citep[CSS,][]{christensen2012catalina,Jedicke2016}, have discovered a large fraction of these objects by covering the northern sky to an apparent magnitude of $m_r \sim$ 21.5-22.0. Unlike the other populations, a substantial fraction ($\sim$40\%) of the 5,000 known TNOs were discovered by three dedicated TNO surveys, the Deep Ecliptic Survey \citep[DES;][]{millis2002deep, adams2014biased}, the Outer Solar System Origins Survey \citep[OSSOS;][]{OSSOS_I} and the Dark Energy Survey \citep[DES;][]{DES}, which traded sky coverage for depth, searching smaller on-sky survey areas to $m_r \sim 23.6-25.2$.

The Vera C. Rubin Observatory's Legacy Survey of Space and Time \citep[LSST;][]{ivezic2019lsst} is expected to provide the next major leap forward in solar system science, expanding the sample of known objects in both depth and breadth. LSST will reach single-image depths around $m_r \sim 24.0$ over 27,000 deg$^2$, including 19,600 deg$^2$ with over 800 repeat visits during its 10 years. By comparison, the coverage of the original Pan-STARRS survey included on average 80 repeat visits of approximately 30,000~deg$^2$ over six years \citep{PS1_surveys}. One of LSST's four main science goals is to ``[Take] an Inventory of the Solar System", discovering as many objects as possible in each orbital class \citep{LSST_book}.  The ten-year survey duration, fast cadence, and 6-filter optical/NIR coverage will also enable detailed characterization of the orbits, sizes, and surface colors for a significant portion of the discovered objects.

It is clear just from Rubin's image depth, sky coverage, and cadence that the system has the potential to dramatically increase the number of known small bodies in the Solar System. Studies of Rubin's performance early in its design phase and mid-way through construction estimated its discovery and characterization yield at around $\sim5$ million bodies across all populations \citep{2007IAUS..236..353I,LSST_book, Jones_Juric_Ivezic_2015, ivezic2019lsst}. However, a full-scale simulation of LSST using a recent model solar system, the near-final observing cadence, the close-to-as-built system performance (including the modeled effects of software performance) has never been performed. This work fills that gap. Using \texttt{Sorcha} \citep{Sorcha, Sorcha_Matt}, a new high-fidelity Solar System survey simulator, we perform the most accurate simulation of Rubin's performance in solar system discovery and quantify its expected discovery and characterization yields on four major solar system populations. These simulations give detailed, quantitative, insight into what LSST will deliver, and allow one to understand (and prepare for) the science opportunities this dataset will enable.

In Section \ref{section:methods} we describe our methods: the simulation technique and code, adopted Rubin observatory parameters and observing cadence, and the small body population models used to simulate LSST discoveries. This is followed by sections presenting expected discovery yields of individual populations. In Section \ref{section:NEOs}, we provide a detailed update to the NEO population estimates of \cite{LSST_book}, \cite{verevs2017high}, \cite{JONES2018181}, and \cite{ivezic2019lsst}, including additional results concerning the expected orbital and size distributions of discovered objects, per-band distributions of detections per object, color and lightcurve characterization, at what point in the survey objects will be discovered, and their end-of-survey observation arcs. Sections \ref{section:MBAs}, \ref{section:Trojans}, and \ref{section:TNOs} do the same for the MBAs, Jupiter Trojans, and Trans-Neptunian objects. In Section \ref{section:early} we discuss what will be possible with two years of LSST data, and we conclude in Section \ref{section:conclusion} with a summary of major results. A companion paper, \cite{Sorcha_Joe}, presents similar predictions for the Centaur population.

\section{Methodology and Inputs} \label{section:methods}

\subsection{Survey Simulator: Sorcha} \label{section:sorcha_settings}
We simulate LSST's small body discoveries using \texttt{Sorcha} \citep{Sorcha, Sorcha_Matt}, an open-source, modular, catalog-level solar system survey simulator designed and built to enable studies like this. Given a description of an observatory (principally the location, camera footprint, and filters) and a simulation of its survey (a database of exposures to be taken), \texttt{Sorcha} precisely integrates the orbits of input bodies, evaluates their detectability in each exposure, and generates a catalog of simulated astrometric and photometric measurements and uncertainties. Then, given a model of the observatory's discovery pipeline methodology and efficiency, it computes whether there are sufficient observations to unambiguously link an object and determine its orbit -- its discovery status. The resulting dataset allows us to assess the number of discoveries in each population (and their properties), as well as evaluate per-object metrics such as the number of detections in each band, observation arcs, and lightcurve and color constraints. A detailed description of \texttt{Sorcha} is given in a separate paper, \cite{Sorcha}, with a full list of simulated quantities listed in Table~6 of that paper. 

For the Rubin Observatory, we adopt values based on a combination of measurements from as-built components and present-day estimates of operational performance, with the most consequential inputs and choices explained in the subsections to follow. The parameters values described here are also given in the format of a \texttt{Sorcha} config file in the repository at \href{https://www.canfar.net/citation/landing?doi=25.0062}{https://www.canfar.net/citation/landing?doi=25.0062}. For further discussion of these parameters as baseline expectations for LSST, we refer the reader to \cite{Sorcha}. 

\subsection{Input Populations}

To estimate the discovery and observation yields for each population, we adopt and extend recent literature models for the well-constrained major populations: NEOs, MBAs, Jupiter Trojans, and TNOs. We do not attempt to model scientifically interesting small populations like Sednoids or Earth impactors. Each model population is discussed in the following subsections and summarized in Table \ref{table:input_pops}. For all populations we assume objects have band-independent phase curves, are not variable (e.g., due to shape or rotation), and exhibit no cometary activity. Since our models consist of identically and independently drawn objects, the sample uncertainty of our population estimates is equal to the square root of that number, except in the case of upscaling where we explicitly list the uncertainty. Input orbit catalogs for all populations are available in the repository at \href{https://www.canfar.net/citation/landing?doi=25.0062}{https://www.canfar.net/citation/landing?doi=25.0062}.

\begin{deluxetable*}{c|cccc}[ht!]
	\tabletypesize{\footnotesize}
	\tablecaption{Input Population Overviews}
	\tablehead{\colhead{Component} &  \colhead{NEO} &  \colhead{MBA} & \colhead{Jupiter Trojans} & \colhead{TNO}}
	\startdata
 Orbits & \texttt{NEOMOD3} \citep{NEOMOD3} & S3M \citep{grav2011} & \cite{TROJANMOD} & CFEPS-L7 \citep{Petit2011} \\ 
 Size Distribution & \texttt{NEOMOD3} & 80\% S3M \citep{wagg2024} & \cite{TROJANMOD} & OSSOS, see Table \ref{table:tno_input} \\
 Colors & C + S & C + S & \cite{wong2015trojans}, \cite{DES_Trojans} &  Red + Blue\\
 Color Fraction & \cite{Morbidelli2020} & \cite{schwamb2023} & \cite{wong2015trojans} & Col-OSSOS, see Table \ref{table:tno_input} \\ 
 Simulated Objects & $4.29 \times 10^8$ & $1.11 \times 10^7$ & $7.07 \times 10^5$ & $3.3 \times 10^6$ \\  
 Size Range & $d >1\, \mathrm{m} $ & $H_r <$ 26.4 & $H_r <$ 18.75 & See Table \ref{table:tno_input} \\
	\enddata
\tablecomments{Summary information for input populations. Note that our NEO model, the only population to include a debiased albedo distribution, is parametrized by diameter while other populations are parametrized by absolute magnitudes. The TNO model is broken up into nine dynamical subpopulations further described in Table \ref{table:tno_input}.}
\label{table:input_pops}
\end{deluxetable*}

\subsubsection{Near Earth Objects} \label{section:neo_pop}
Our NEO input population is a derivative of \texttt{NEOMOD3} \citep{NEOMOD3}, a state-of-the-art debiased NEO population model which provides objects' size, orbit, and albedo. We use the \texttt{NEOMOD3} object generator to generate a population of orbits with semi-major axis ($a$), eccentricity ($e$), and inclination ($i$), as well as absolute magnitudes ($H_V$), diameters ($d$), and albedos ($p_V$), to which we only add the orbital angles $\omega$, $\Omega$, and $M$ and colors. We assign orbital angles uniformly from $[0, 2\pi)$. Colors are assigned from the two most common NEO spectral types, S and C, according to a criterion from \cite{Morbidelli2020}: objects with $p_V \geq 0.1$ are assigned S-type, while objects with $p_V \leq 0.1$ are assigned C-type. Each object is given the colors of a random reference object with its corresponding spectral type (see Section \ref{section:colors}). For simplicity, we use the IAU standard $HG$ phase curve model \citep{bowell1989}, with a constant assumed phase slope $G = 0.15$ for all bands and objects, matching the Minor Planet Center assumption for unconstrained slope parameters\footnote{\href{https://minorplanetcenter.net/web_service/}{https://minorplanetcenter.net/web\_service/}}. 

We produce a full-scale 6.1 million-object model of the population with diameter $d \ge 10 \, \mathrm{m}$. Due to computational constraints, we could only simulate 428 million of the full-scale 1.9-billion-object population with diameter $1$--$10 \, \mathrm{m}$. We weight this population by a factor of 4.42 (i.e. we report numbers of objects as 4.42 times larger) to produce a full-scale NEO population down to diameters of 1 meter. When discussing NEOs by size, we use this convenient diameter cutoff of $10 \, \mathrm{m}$ for the smallest category of NEOs, and the traditional planetary defense cutoff of $140 \, \mathrm{m}$ \citep{USC_16691} for the largest.

\subsubsection{Main Belt Asteroids} \label{section:mba_pop}
Our MBA input population is based on the Pan-STARRS Synthetic Solar System Model \cite[S3M,][]{grav2011}, which includes a full-scale population of objects with six-dimensional orbits and $H_V$ magnitudes, leaving us to generate colors and phase slope parameters. MBAs are assigned to S-type with probability $a/2 \mathrm{au}-1$, following \cite{schwamb2023}, and otherwise C-type. Then, each object is uniformly assigned the colors of one reference object of its spectral type, as described in Section \ref{section:colors}. As in the NEO case, MBAs are given an $HG$ phase function with a slope of $G = 0.15$ matching the Minor Planet Center default.

S3M was calibrated to a relatively bright and small sample of objects and was designed for a survey shallower than LSST, causing two issues. First, in a comparison of S3M to modern observations, \cite{wagg2024} suggests that an 80\%-scale S3M better fits the modern number of $m \sim 20$ detections of asteroids. We adopt this modification, discard a uniformly drawn 20\% of the original model, and are left with 11.1 million MBAs to simulate.

\subsubsection{Jupiter Trojans} \label{section:tr_model}

Our Jupiter Trojan input population is an extrapolation of the debiased orbit-magnitude model of \cite{TROJANMOD}, which separately models the Trojans orbiting Jupiter's L4 and L5 points. While the original model is well-calibrated to $H_V \sim 15$, we extend the population to $H_V \sim 19$ using the original L4 population's small-end slope to avoid the problem of a divergent L4/L5 ratio. We assign two color classes as in \cite{wong2015trojans}, using their mean $V-r$ and magnitude-dependent $g-i$ and color fraction, while a mean $r-i$ and $r-z$ are taken from \cite{DES_Trojans}, and $u-r$ and $r-y$ colors are the average D-type colors described in Section \ref{section:colors}. Again, a slope of $G = 0.15$ is uniformly assigned.

\subsubsection{Trans-Neptunian Objects} \label{section:tno_pop}
\begin{deluxetable*}{c|ccc|cc}[ht!]    
    \tablecaption{Outer Solar System populations}
	\tabletypesize{\footnotesize}
	\tablehead{\colhead{Population} &  \colhead{$H_r$ distribution} &  \colhead{Normalization} &\colhead{Reference} & \colhead{Blue Fraction\tablenotemark{a}} & \colhead{Max Depth ($H_r$)}}
	
	\startdata
	 Cold Classicals & Tapered exponential\tablenotemark{b} & 11,000 ($H_r < 8.3$) & \cite{kavelaars2021} & 0.000 & 9.79\\ 
	 Hot Classicals & Tapered exponential\tablenotemark{b}  & 20,000 ($H_r < 8.3$) & \cite{petit2023} & 0.775 & 10.25\\
	 Detached & Single exponential\tablenotemark{c} & 36,000 ($H_r < 8.66$) & \cite{beaudoin2023} & 0.775 & 10.71\\
	 Scattering & Divot\tablenotemark{d} & 90,000 ($H_r < 8.66$) & \cite{Lawler2018a} & 0.775 & 16.38\\
	 Resonant (3:2) & Single exponential\tablenotemark{e} & 8,000 ($H_r < 8.66$) & \cite{Volk2016} & 0.869 & 11.62 \\
	 Resonant (7:4) & Single exponential\tablenotemark{e} & 1,000 ($H_r < 8.66$) & \cite{Volk2016} & 0.645 & 10.73\\
	 Resonant (2:1) & Single exponential\tablenotemark{e} & 5,200 ($H_r < 8.66$) & \cite{Volk2016}  & 0.763 & 11.00\\
	 Resonant (7:3) & Single exponential\tablenotemark{e} & 4,000 ($H_r < 8.66$) & \cite{Volk2016}  & 0.847 & 11.35\\
	 Resonant (5:2) & Single exponential\tablenotemark{e} & 5,700 ($H_r < 8.66$) & \cite{Volk2016} & 0.847  & 12.15\\ 
	\enddata
    \tablenotetext{a}{From \cite{pike2023} except Cold Classicals (see text)}
    \tablenotetext{b}{Equation \ref{eq:tapered}, $\alpha = 0.4$, $\beta_{\rm SI} = 0.42$, $H_{\rm B} = 8.1$}
    \tablenotetext{c}{Equation \ref{eq:singleplaw}, $\alpha =0.6$ }
    \tablenotetext{d}{Equation \ref{eq:divot}, $\alpha_{\rm b} = 0.9$, $\alpha_{\rm f} = 0.3$, $H_{\rm B} = 8.3$, $c = 3.2$}
    \tablenotetext{e}{Equation \ref{eq:singleplaw}, $\alpha = 0.9$}
\label{table:tno_input}    
\end{deluxetable*}

Our TNO model combines the Canada-France Ecliptic Plane Survey's \citep{Jones2006} L7 orbital model \citep[CFEPS-L7;][]{Petit2011} along with absolute magnitude and color fractions from subsequent surveys to generate an outer Solar System model comprised of nine subpopulations: the hot and cold classical components, the detached and scattering excited components, as well as the populations in the 3:2, 7:4, 2:1, 5:3, and 7:3 mean resonances with Neptune. While this is a simplification of the complex dynamical behavior seen in the outer Solar System, these populations represent the majority of objects that have been discovered in previous surveys \citep[\emph{e.g.,}][]{Petit2011,OSSOS_VII,DES,smotherman2024}. A full description of each subpopulation is given in Table \ref{table:tno_input}. 

For each subpopulation, we draw absolute magnitudes from a reference distribution down to a limit defined as where the lowest-perihelion object would have an $r$-band apparent magnitude of 25.5, corresponding to LSST's best-case $r$-band limiting (24.5) magnitude, plus a 1-magnitude buffer to account for other colors and low-probability sub-limit detections. After orbits are assigned, our per-population magnitude cut leaves the non-classical populations dominated by objects which are undetectable due to their distant perihelia and high absolute magnitudes. We repeat the magnitude-at-perihelion cut on a per-object basis for these subpopulations, eliminating the bulk of undetectable objects and reducing our computational load.

Before drawing each object's $(a, e, i)$ vector from the CFEPS-L7 objects in its subpopulation, we re-classify certain CFEPS subpopulations to match the common definitions. First, we limit the ``hot classical'' orbits to those with $a$ below the 2:1 resonance, as in \cite{petit2023}. We combine those with greater $a$ with those labeled as ``scattering'' to form the ``scattering'' and ``detached'' subpopulations following the classification procedure of \cite{2008ssbn.book...43G,Khain2020} using the implementation of \cite{DES}: their orbits are integrated over 10~Myr with \texttt{REBOUND} \citep{rein2012}, and those whose maximum $a$ variation changes by 3.75\% from the initial value over this period are defined as ``scattering,'' the rest being ``detached.'' We assign uniformly-distributed $\omega$, $\Omega$, and $M$ when we clone non-resonant objects, which are not known to have preferred directions in the Solar System. In contrast, resonant objects' orbital angles are strongly constrained by their resonance with Neptune, so we instead assign clones to the original CFEPS-L7 angles perturbed only by only a uniformly drawn $\pm 1 \degr$. 

We use three different parametrizations for absolute magnitude distributions, corresponding to different behavior seen in TNO subpopulations. The simplest case, which we use for the resonant and detached populations, is that of a single exponential power law, where the cumulative distribution
\begin{equation}
    p(\leq H; \alpha) \propto 10^{\alpha H} \label{eq:singleplaw}
\end{equation}
has a single parameter $\alpha$ corresponding to the power law slope. Following \cite{Lawler2018a}, we posit that the scattering population is modeled by a power law with a divot,
\begin{equation}
    p(\leq H;\alpha_\mathrm{b}, \alpha_{\mathrm{f}}, H_\mathrm{B},c) \propto \begin{cases} 10^{\alpha_\mathrm{b} H}, & H \leq H_\mathrm{B},\\ 10^{\alpha_\mathrm{b} H_\mathrm{B}} +  \frac{1}{c} 10^{\alpha_\mathrm{f} (H - H_\mathrm{B})} & H \geq H_\mathrm{B}
    \end{cases}.
\label{eq:divot}
\end{equation}
Here, $H_\mathrm{B}$ is the magnitude where the factor-of-$c$ divot and transition between the bright-end slope of $\alpha_\mathrm{b}$ and the small-end slope of $\alpha_\mathrm{f}$ occur. Finally, we use the exponential tapered power law of \cite{kavelaars2021}, motivated by simulations of the streaming instability hypothesis of planetesimal formation \citep{youdin2005streaming,simon2024comets}, where the cumulative distribution is given by
\begin{equation}
    p(\leq H; \alpha_\mathrm{SI}, \beta_\mathrm{SI}, H_\mathrm{B}) \propto 10^{\alpha H} \exp\left(- 10^{-\frac{3}{5} \beta_\mathrm{SI} (H_r - H_\mathrm{B})}\right)
\label{eq:tapered}
\end{equation}
As before, $\alpha$ is the slope (here, on the small end, $H \gtrsim H_\mathrm{B}$), $H_\mathrm{B}$ is the transition magnitude, and $\beta_\mathrm{SI}$ is the tapering parameter that controls the bright end of the distribution. In all cases, we sample from each distribution by inverse transform sampling.

We choose a simplified ``red'' and ``blue'' two-component color model, as is commonly done in the literature \citep[e.g.,][]{fraser2012,Peixinho2015,Schwamb2019}, with the colors derived in Section \ref{section:colors}. Each population is given a color fraction following \cite{pike2023}, where we associate their faint infrared class with our red objects, and their bright infrared class with our blue objects (see Table \ref{table:tno_input} for values). We note that, while our cold classical blue fraction of 0 does not include the subset of blue binaries \citep{fraser2017}, these represent a minority of the population, with still poorly constrained fractions \citep[see \emph{e.g.}][]{bernardinelli2025photometry}. Finally, as TNOs are typically only observed at very small phase angles and phase effects are not known for all broadband filters, we ignore these minor and poorly-constrained phase effects and do not assign phase curves to TNOs.

\subsubsection{Colors} \label{section:colors}
Across our populations, we have five spectral classes (C-type, S-type, D-type (TNO red, and TNO blue for which we produce LSST colors. We start with a set of objects in each class with measured optical spectra: 384 S-, 143 C-, and 10 D-type spectra from the Small Main-Belt Asteroid Spectroscopic Survey \citep[Phase II;][]{bus2002a}, 1999~OX$_3$ \citep{1999OX3} as our representative red TNO, and 2002~PN$_{34}$ \citep{2002PN34} as our representative blue TNO. We linearly extrapolate these spectra to the poorly-constrained extremes of LSST wavelengths \citep[320–1050~nm;][]{LSST_book}, multiply by the Solar spectrum \citep{kurucz05} to convert them to spectral energy distributions (SEDs), and then integrate the resulting flux under the corresponding LSST filter bandpasses to produce colors. The mean and standard deviation for each population's colors are shown in Table \ref{table:colors}.

Using constant colors, as we do for the Jupiter Trojans and TNOs, underestimates the width of populations' color distributions and therefore the width of their detection likelihood function. We therefore expect to systemically under-predict detections of dim Jupiter Trojans and TNOs and over-predict detections of bright ones. Previous surveys measuring the effect of color on discovery probability find effects of only a few percent per 0.1 magnitudes \citep{DES}, so we do not expect this to have a large effect on overall discovery or characterization yields.

\begin{deluxetable*}{c|cccccc|c}[ht!]

  	\tablecaption{Model colors\label{table:colors}}

	\tabletypesize{\footnotesize}
	\tablehead{\colhead{Spectral type} &  \colhead{$u-r$} &  \colhead{$g-r$} & \colhead{$r-i$} & \colhead{$r-z$} & \colhead{$r-y$} & \colhead{$V - r$}  &\colhead{Reference} }

	\startdata
 S-type & $1.62 \pm 0.13$ & $0.78 \pm 0.03$ & $0.5 \pm 0.01$ & $0.71 \pm 0.03$ & $0.92 \pm 0.07$ & $0.35 \pm 0.01$ & \cite{bus2002a}\\ 
 C-type & $1.28 \pm 0.11$ & $0.62 \pm 0.02$ & $0.43 \pm 0.01$ & $0.72 \pm 0.02$ & $0.93 \pm 0.07$ & $0.27 \pm 0.01$ & \cite{bus2002a}\\ \hline
 Trojan & 1.44 & 0.51, 0.64 & 0.22 & 0.39 & 1.18 & 0.25 & See section \ref{section:tr_model} \\  
 TNO Red & $2.73$ & $0.92$ & $0.44$ & $0.67$ & $0.84$ & - & \cite{1999OX3}\\ 
 TNO Blue & $1.95$ &  $0.62$ & $0.27$ & $0.38$ & $0.46$ & - & \cite{2002PN34}\\  
\enddata
 
\tablecomments{Colors for each spectral class. For spectral classes with multiple representatives, standard deviations are given. V represents the Johnson $V$ band, while \textit{ugrizy} are from the LSST photometric system.}
\end{deluxetable*}

\begin{figure*}
\begin{center}
\includegraphics[width=\textwidth]{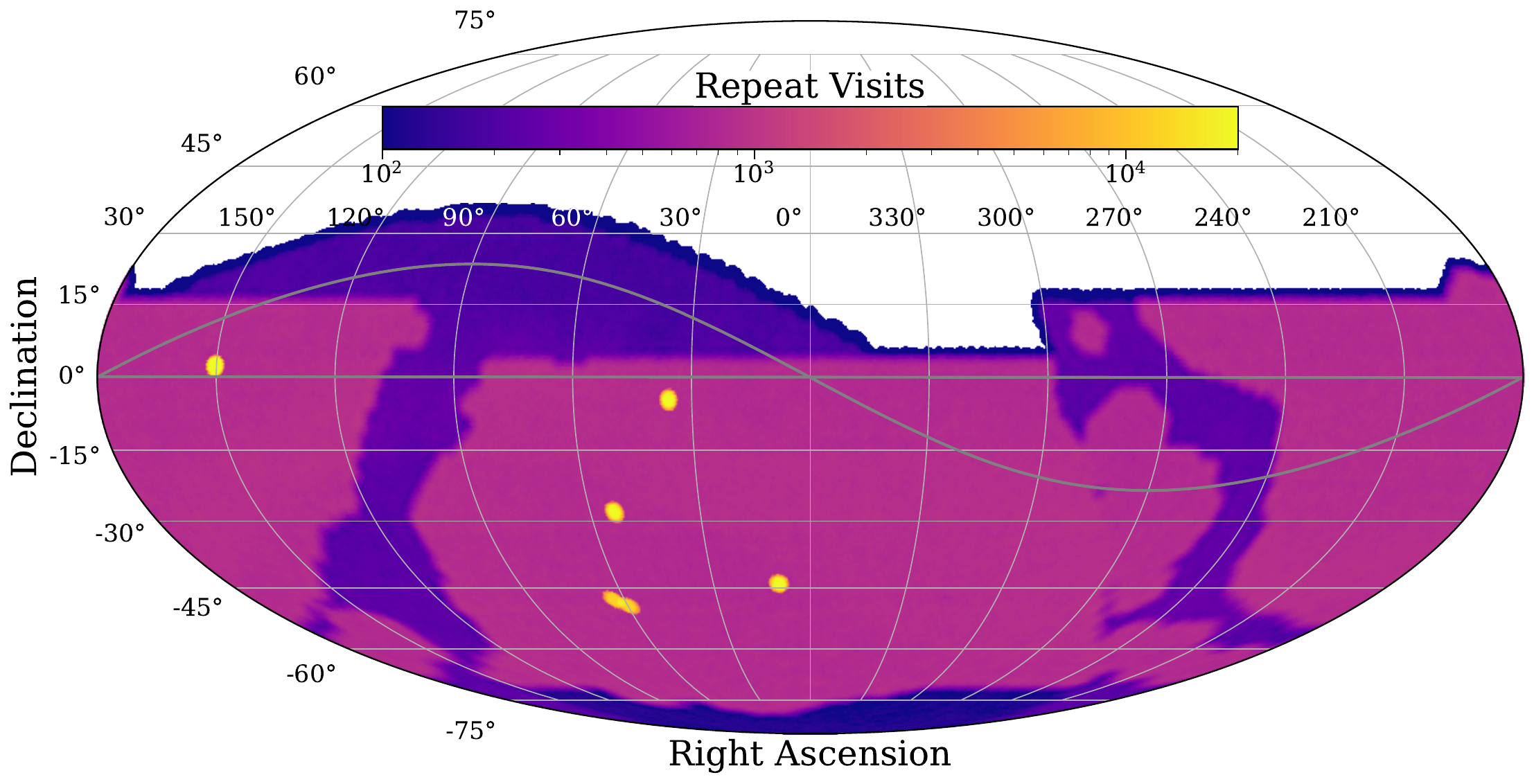}
\end{center}
\caption{A heatmap of the sky coverage of the LSST baseline v3.4 cadence \citep{baseline_v3.4}, at the completion of the 10-year LSST survey. The main survey area -- the so-called ``wide, fast, deep`` (WFD) region -- results in a relatively uniform number of visits over approximately $19,600$deg$^2$ of the sky shown in magenta. The DDFs fields, having $>10,000$ of revisits are shown in yellow. The mini-surveys (including the northern Ecliptic spur, the Galactic plane, and the southern polar cap) generally have fewer visits than the WFD area, and are shown in purple. Note that a $\sim 15{\rm deg}$ band around the ecliptic is covered in full to maximize LSST's solar system discovery and science potential.} \label{fig:full_survey}
\end{figure*}

\subsection{LSST survey pointing simulation} \label{section:cadence}

Our simulations assume the ``baseline'' simulation of the LSST v3.4 cadence \citep{baseline_v3.4} as our simulated LSST pointing database. Generated by \texttt{rubin\_sim} \citep{bianco22, yoachim23} and the Rubin Observatory scheduler, \texttt{rubin\_scheduler} \citep{naghib19, yoachim24b}, the v3.4 baseline simulation represented the best available approximation of the LSST exposure dataset as of our analysis\footnote{Version 4.0 has been released since the completion of this work incorporating minor tweaks and fixes to v3.4 (in spite of apparently large change in version number). Optimization and new versions will continue to be produced until the survey begins, but large qualitative changes are not likely.}. The baseline cadence consists of two million pointings, each with associated central sky coordinates, detector rotation, reference epoch, exposure duration, and filter, as well as airmass and sky brightness and the resulting limiting magnitude of each exposure. In the baseline simulation, LSST starts on 2025 May 1 and runs continuously for 10 years. Most of its observing time is spent on the wide-fast-deep (WFD) survey, with a cadence that covers the visible southern sky with a pair of 30-second visits about 30 minutes apart every three or four days, distributed between the \textit{ugrizy} filters \citep{jones2021survey}. In addition to the WFD survey, five ``deep drilling`` fields (DDFs) will be surveyed with a much higher cadence. Assorted mini- and micro-surveys will cover an additional $\sim7,500$ deg$^2$ with lower cadence, including the Dusty (Galactic) Plane, the South Celestial Pole, the Northern Ecliptic Spur (NES), which completes the survey's coverage of the ecliptic plane in $\textit{griz}$ but excludes $\textit{u}$ and $\textit{y}$, and the twilight NEO survey, which observes at low elevation towards the Sun to observe objects within the Earth's orbit. Figure \ref{fig:full_survey} shows a map of the sky coverage of v3.4 cadence, while \cite{PSTN-055} and \cite{schwamb2023} contain more detailed explanations of the survey strategy and its optimization for solar system discovery.

\subsection{Detection and Discovery Efficiency} \label{section:discovery}
\texttt{Sorcha} models LSST’s production of 5$\sigma$ single-exposure difference image source catalogs\footnote{At present, we do not simulate discoveries enabled by techniques such as shift-and-stack \citep[e.g.][]{Smotherman_2021}.}. The list of objects in each simulated visit is constructed by matching the on-sky position of each object against a high-precision model of the LSST camera footprint, including in particular the gaps between CCD chips. The probability of detection of each source is modeled by a logistic function with width parameter 0.1 and 50\% detection probability at the exposure's limiting magnitude, matching \cite{verevs2017high}. Detections brighter than LSST's bright limit of of 16.0~mag are filtered out, since saturated sources will not be well-measured\footnote{We do not precisely model the bright end, as there's still considerable uncertainty -- on order of $0.5~{\rm mag}$ -- about exactly where Rubin's detectors will saturate. Also, most of the discovery potential is likely on the faint end}. These constant source detection parameters may underestimate the breadth of the detection probability function when image quality is very low, underestimating the number of detections contributed by poor-quality images. We do not model the difficulties of source detection in high-density regions such as the galactic plane or globular clusters, which could prevent asteroid detections where templates cannot be produced.

LSST's Solar System linking pipeline is modeled as matching its design requirements: discovering 95\% of objects detected at least twice per night on at least three nights within a window of 14 days. Objects which have multiple sets of three-night observations are given multiple independent 95\% chances at discovery. \cite{kurlander2025well} discusses issues with debiasing a survey which uses HelioLinC as its discovery algorithm, but until the pipelines' efficiency is actually measured, a flat 95\% probability per discovery chance is used. We require a tracklet to have length at least 0.5 arcseconds (2.5 pixels) to be usable for solar system linking, corresponding to the apparent motion of a stationary object 150$\mathrm{au}$ away at opposition. 

Once discovered, we assume the survey will perfectly pre-cover any prior detections of linked objects so that all of their detections are recovered in the catalogs regardless of orbital uncertainty at their time of discovery. This is a reasonable assumption given most objects observed in LSST have very long arc (e.g. see Table~\ref{table:discovery_summary}).

\subsection{Color and Light Curve Metrics}
\label{section:metrics}

Besides astrometry and linking, Rubin pipelines will attempt to characterize the physical properties of observed small bodies. These include the derivation of absolute magnitude and slope parameters (in each observed band), as well as the rotational periods and light curves.

To evaluate whether an object's set of observations provides sufficient data to extract high-quality colors and lightcurves, we use metrics described in \cite{schwamb2023} and available in the Rubin Observatory Metrics Analysis Framework \citep[MAF; ][]{MAF}. As discussed in \cite{dobson2023phase}, these metrics quantify whether it is possible to perform very high-quality color and lightcurve measurements: the color metric proxies whether data will be sufficient to fit per-filter phase curves and simple lightcurves to account for objects' features in each filter, while the lightcurve metric proxies whether the generation of a shape model will be possible from lightcurve data. Statistical studies of colors and lightcurves are often performed with fewer or less precise data points \citep[see, for example,][]{durech2020,schemel2021,bernardinelli2023}, making our predictions of the size of color and lightcurve catalogs conservative. More colors and lightcurves than we predict be extracted, higher uncertainties can be tolerated. 

The lightcurve metric counts the number of objects whose number and quality of observations is deemed sufficient to produce a high-quality folded light curve. The criterion for this metric is based on an object's range of phase angle, range of ecliptic longitude, and the largest number of signal-to-noise (SNR)-weighted detections available in any one band. To qualify, an object must be observed over at least a $5 \degr$ phase angle range and a $90\degr$ range of ecliptic longitude. Since distant objects cannot move in phase angle or ecliptic longitude over 10-year timescales, we drop these two requirements for TNOs and accept that TNO lightcurves will not well-constrain phase curves. Though it is described slightly differently in \cite{schwamb2023} and \cite{robinson2024tuninglegacysurveyspace}, the SNR threshold is exactly equivalent to requiring that the sum of SNR of all detections in one band is at least 5,000, truncating the contribution from any single detection at 100 if it is greater. This metric is described in Table \ref{table:metrics}.

The color metric is based on a threshold of SNR-weighted detections available in the \textit{griz} bands. To meet this metric, objects must have an SNR sum in each \textit{griz} band of at least 200 with the maximum contribution from single detections truncated at 20. The MAF color metric for TNOs is slightly different -- a primary band with 30 detections and 3 other bands with 20.

\begin{deluxetable*}{cc|cccc}[ht!]
\label{table:metrics}

	\tabletypesize{\footnotesize}
	\tablecaption{MAF / \cite{schwamb2023} metric parameters}
	\tablehead{\colhead{Component} & \colhead{Metric} & \colhead{Bands}  & \colhead{SNR sum} & \colhead{Threshold} & \colhead{Additional Requirements}}
	\startdata
Non-TNO & Light Curve & Any one & 5,000 & 100 & $5$ deg phase, $90$ deg ecliptic longitude \\ 
TNO & Light Curve & Any one & 5,000 & 100 & - \\ 
Non-TNO & Colors & \textit{griz} & 200 & 20 & - \\ 
TNO & Colors & \textit{griz} & 100 & 5 & Primary band with 150 SNR sum\\ 
    \enddata
\tablecomments{Parameters of the light-curve and color MAF metrics used to count the number of objects suitable for high-quality light curve and color determination (see text for details).}
\end{deluxetable*}

\section{The Simulated LSST Solar System Object and Detection Catalog} \label{section:simulation}

\subsection{Overview}

We ran our {\tt Sorcha} simulations on two high-performance computing clusters: Odyssey at Harvard University and Bura at the University of Rijeka. Input catalogs were split into batches of 48,000 objects and each processed by 12 cores per node, typically taking 30-50 minutes per batch to simulate the 10-year survey. The simulation time was dominated by the very large number of small NEOs; around 60,000 core-hours were spent simulating the NEOs, with 2,350 spent on the MBAs, 168 on the Trojans, and 630 on the TNOs. This is roughly equivalent to a month of continuous computing on a modern 100-core machine.

The process generated a catalog of 1.145 billion detections of solar system objects collected over ten years of simulated observing. This catalog of observations -- our simulated LSST -- is more than twice as large as the $4.762 \times 10^8$ detections listed in the MPC observations database as of Jan 22, 2025. Comparing average rates, we find that LSST will contribute about 110 million observations per year (or about 300,000 per day), 2.5 times more than the 44.7 million detections (122,000 per day) submitted to the MPC by all surveys combined in 2024. 

\begin{figure*}
\begin{center}
\includegraphics[width=\textwidth]{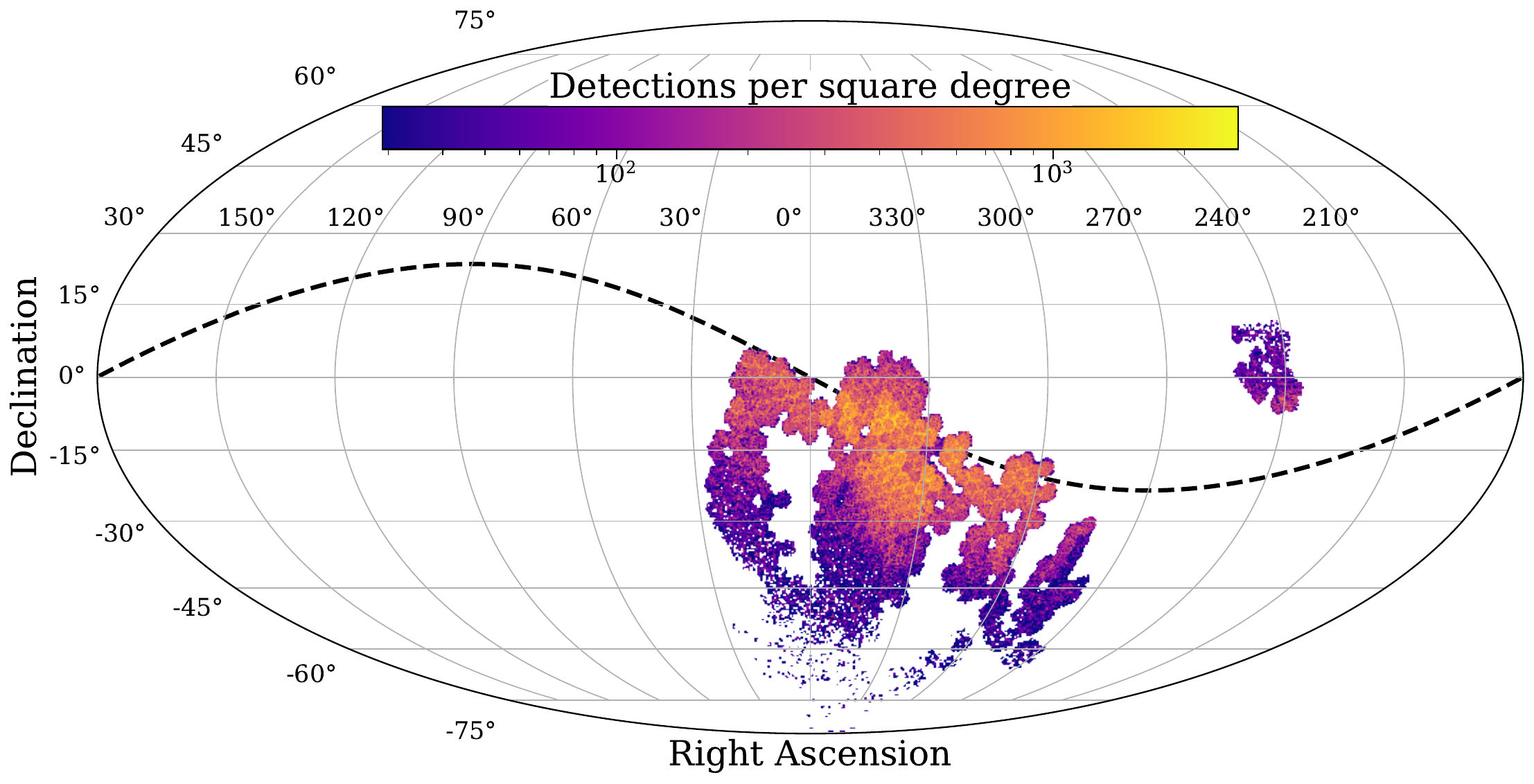}
\end{center}
\caption{A heatmap of the simulated detections from a typical night (2025-08-17), with the ecliptic plane dashed in black. The night is spent observing a mostly-contiguous section of the WFD survey, with no spent on the deep drilling fields or mini-surveys. As expected, detections are much more numerous close to the ecliptic.} \label{fig:one_night}
\end{figure*}

In Figure \ref{fig:one_night}, we show the detections collected on a fairly typical night, to illustrate the expected nightly detection numbers as well as the general observing pattern. As expected, the observations of minor planets are the most numerous around the ecliptic, with some NEOs and TNOs found at higher ecliptic latitudes. LSST pointings (the ``visits'') are generally arranged in contiguous zones to minimize slewing, and are mostly within the southern WFD footprint. For this particular night, no time is spent in the NES or DDFs. Rubin's default observing cadence takes a pair of observations of each pointing, allowing for the generation of two-observation tracklets and their subsequent linking over 14-day windows. There are slight overlaps between adjacent fields, which leave a visible grid pattern in counts of observations. These locations -- about $\sim 10$\% of the survey area -- are observed up to four times per night, enabling the production of four-detection tracklets for sufficiently slow-moving objects that happen to be there.

\begin{figure*}
\begin{center}
\includegraphics[width=0.8 \textwidth]{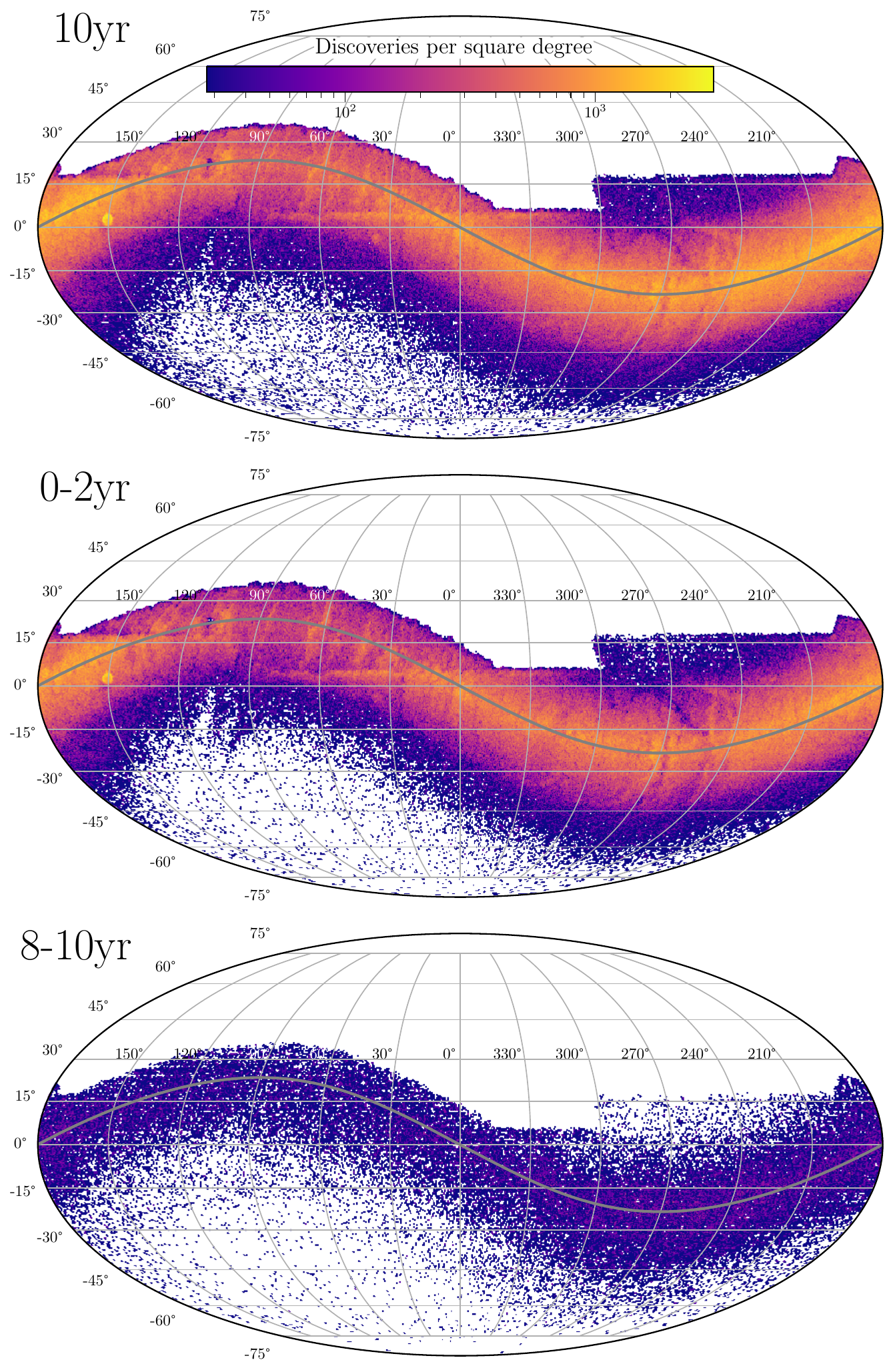}
\end{center}
\caption{Heatmap of the (equatorial) on-sky positions of discovered objects over the full survey (top panel), first two years (second panel), and final two years (bottom panel). Discoveries are concentrated on the ecliptic plane. The discoveries in the first two years comprise a large fraction of the full survey's discoveries, though late-survey discoveries are still substantial. Bright objects in the NES which happen to not be discovered early in the survey are often discovered as they enter the WFD survey area, leading to an overdensity of objects at the western NES-WFD boundary and a underdensity at the eastern boundary.} \label{fig:discovery_map}
\end{figure*}

\begin{figure}
\begin{center}
\includegraphics[width=0.49 \textwidth]{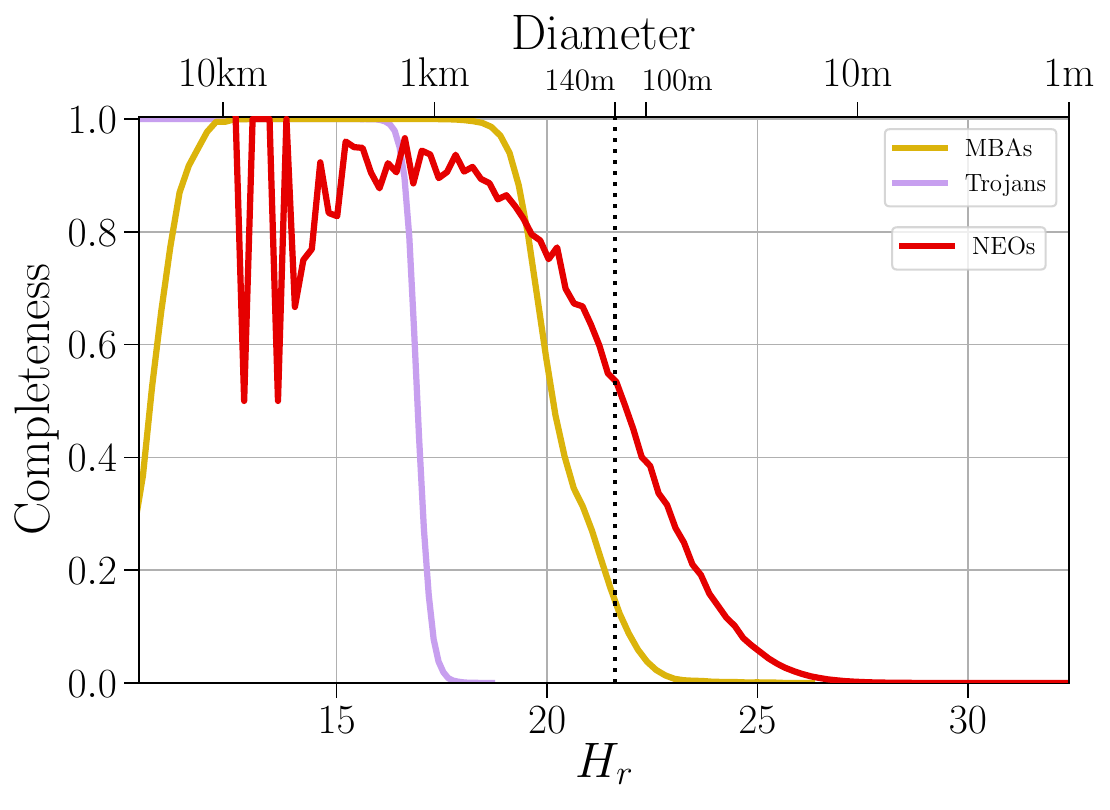}
\end{center}
\caption{
Fraction of simulated objects discovered (``discovery completeness'') for NEOs, MBAs and Jupiter Trojans. The NEO population is measured in diameter while the MBAs and Trojans are measured in $H_r$. Bright-end loss of completeness is due to bright source saturation. The $H_r$ and diameter axes are aligned assuming a reference albedo of 0.25. The Jupiter Trojans, being relatively spatially confined, drop sharply from very high to very low completeness over a small range of absolute magnitudes. NEO completeness shrinks, but does not reach zero even at diameters of 1~$\mathrm{m}$.}\label{fig:total_completeness}
\end{figure}

In our simulated LSST, there are 2,809 nights such as the one in Figure~\ref{fig:discovery_map}. Applying Rubin software's linking and discovery criteria, we find they allow us to discover 5,356,423 objects over the 10 years of the survey. These include 1.27E5 NEOs, 5.09E6 MBAs, 1.09E5 Jupiter Trojans, and 3.70E4 TNOs (see Table \ref{table:discovery_summary}). LSST's discovery completeness on these intrinsic populations as a function of size is shown in Figure \ref{fig:total_completeness}. These numbers represent discoveries Rubin would make if no minor planets were known before LSST. Assuming that all 1.4 million presently known objects \citep[retrieved 2025 Jan 31 from the SBDB;][]{JPL_tools} are among these, the LSST will measure properties of some 3.9 million new small bodies, a 3.6x increase over the current number.

The discovery rate as a function of survey time is not constant. The top panel in Figure \ref{fig:discovery_map} shows the on-sky distribution of discoveries for the full 10-year duration survey. The middle and bottom panels show the discoveries made in the first two and the last two years, respectively. A few details stand out. As expected, most objects are discovered within ten degrees of the ecliptic, dominated the MBAs which congregate there. Next, the southern portion of the ecliptic is moderately denser with discoveries than the NES area, reflecting fewer observations Rubin will make in the NES region. Still, observing the NES is extremely important, as 17\% of Rubin discoveries are made in that region, including 11\% of TNOs. We also spot some bright circles, most notably the one around $(\alpha, \delta) \sim (150^\circ, 2^\circ)$. This is a deep drilling field, observed early in the survey and therefore a source of a large number of early discoveries. There are $6$ DDFs spread across the sky; the others are less prominent in this figure because they lie at high ecliptic latitudes. The WFD-NES boundary on the ecliptic is highlighted by an overdensity of inner solar system object discoveries. Secondly, there are some ``vertical'' whispy patterns in the discoveries, resembling waves roughly perpendicular to the line of the ecliptic. These are due to a combination of weather and nightly observing patterns early in the survey, and aren't especially meaningful.

We find that most objects are discovered quite early in the ten-year survey, with distant populations quickly reaching high completeness (see Figure \ref{fig:dates}). Over its first two years (by DR3), LSST will discover $\sim 70\%$ of the objects it will discover over its full ten years, including 72\% of TNOs, 68\% of Jupiter Trojans, 69\% of MBAs, and 53\% of $d >$140$\, \mathrm{m}$ NEOs. With the discovery of a large fraction of objects, accurate population estimates will be possible quickly, contingent only on survey characterization for debiasing. While we do not model Sednoids, Planet 9, or other extreme TNO ($a > 250\, \mathrm{au}$, $q>37\,\mathrm{au}$) populations here, they will be discovered early with the other TNOs, allowing for an early re-evaluation of the evidence for the Planet 9 hypothesis \citep{Shankman2017,Bernardinelli2020,Napier2021,brown2021,Siraj2025}. 

More revealing is what comparing the three panels tells us: that most discoveries occur fairly early in the survey, with relatively few new objects being discovered by years 9 and 10. Quantitatively, 70\% of objects are found within the first two years, with the fraction being larger the more distant the population (i.e. 72\% for TNOs, 73\% for Jupiter Trojans, 69\% of MBAs, and 53\% of $d >140\mathrm{m}$ NEOs), bringing opportunities for early science (see Section \ref{section:early}). This is caused by the depth and reach of Rubin: as it sweeps through the Solar System, within one synodic period it discovers most objects that are brighter than its limiting magnitude over a $\sim 15$day linking window. The subsequent years then fill the gaps caused by weather and pick up additional objects which have moved towards perihelion and/or are at the limit of detectability. In contrast, populations with a relatively constant flux of new objects include the small NEOs (see Section \ref{section:NEOs}) and interstellar objects (not discussed here).

As we have noted, our simulated catalog is comprised of 1.145 billion single-epoch detections which enable a total of 5,356,423 objects to be discovered. Of the 1.145 billion, some 43 million ($\sim 4$\%) belong to objects which are not observed in a pattern suitable for successful linking. For example, an object with a total of just five observations would not be linkable. An object with fifty observations spread evenly across the 10 years would also escape detection. Such observations will be in the LSST source catalog, but won't be recognized as belonging to moving objects. They may be linked outside of regular LSST processing by another  algorithm or precovered when objects are discovered later by other surveys. Still, the inverse is possibly more impressive -- these results imply that some $96$\% of all minor planet observations in the LSST will be linked, characterized, and have an orbit computed. In other words, the LSST will be a highly efficient small body discovery machine.
\\

Having summarized the general properties of the simulated Rubin catalog, we next examine the individual populations in some more depth. We stress that our analysis is far from exhaustive: numerous individual papers could be written for various aspects of each population, that we only briefly discuss below. Rather, our discussion is only meant to summarize the key properties of the simulated sample, and to serve as an illustration of the kinds of analyses the readers are invited to perform themselves using our dataset.

\begin{deluxetable*}{cc|ccccc}[ht!]
\label{table:discovery_summary}

	\tabletypesize{\footnotesize}
	\tablecaption{Summary of LSST solar system catalog properties}
	\tablehead{\colhead{Component} & \colhead{Currently Known} & \colhead{Objects observed} & \colhead{Median Arc} & \colhead{Median Number of Detections} & \colhead{High-quality colors}  & \colhead{High-quality light curves}}
	\startdata
NEOs & 37,932 & $127,040 \pm 557$ & 96~d& 23 & $4,418 \pm 66$ (3.5\%) & $471 \pm 22$ (0.3\%) \\
MBAs &  1,380,217 & $5,087,541 \pm 1,661$ & 9.0~yr & 160 & $1,666,184 \pm 1291$ (32.8\%) & $421,365 \pm 649$ (8.3\%) \\
Jupiter Trojans & 15,134 & $109,367 \pm 331$ & 9.0~yr & 193 & $45,221 \pm 213$ (41.3\%) & $5,846 \pm 76$ (5.3\%) \\
TNOs &  5,246 & $37,002 \pm 192$ & 9.5~yr & 234 & $16,651 \pm 129$ (45.0\%) & 1,057 $\pm 213$ (2.9\%) \\
\enddata
\tablecomments{Currently-Known quantities retrieved from 2025 Feb 24 from the SBDB \citep{JPL_tools}. Provided one-sigma sample uncertainties are the square root of the population size, since our populations made of objects drawn independently and identically. Sample uncertainty contributed by the upscaled small NEO population is upscaled by the same factor.}
\end{deluxetable*}

\subsection{Near Earth Objects} \label{section:NEOs}

\subsubsection{Discovery Yield and Completeness}

\begin{figure}
\begin{center}
\includegraphics[width=0.49\textwidth]{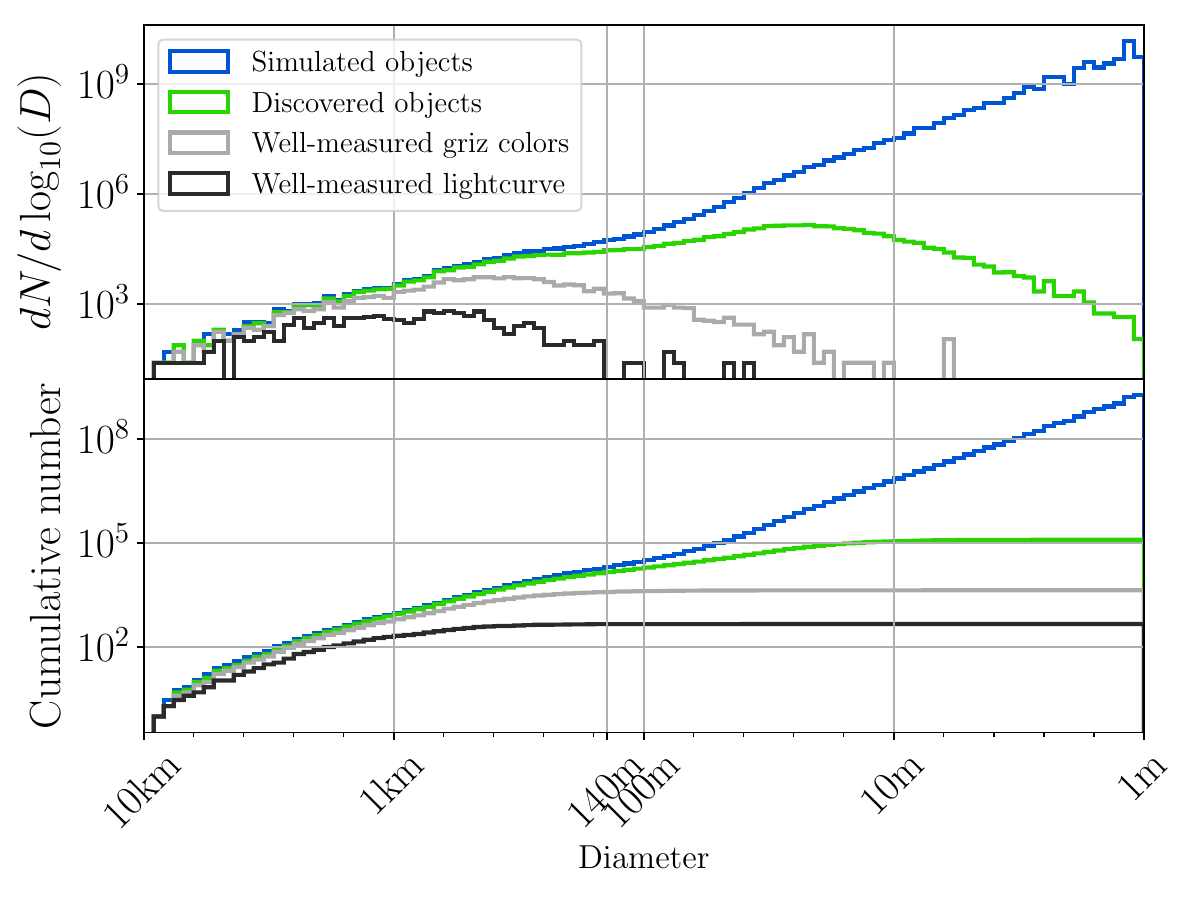}
\end{center}
\caption{The differential number density and cumulative number of NEOs (top and bottom panels respectively). The size distribution of the input population is shown in blue, the discovered sample is shown in green, the subsample with well-measured $griz$ colors is in black, and the well-measured light curve subsample is given in gray. \texttt{NEOMOD3} sizes are given in increments of 0.1~m, which results in a spiky histogram of very small NEOs, even with very large sample sizes. Note that discovered objects are a subset of simulated objects, and well-measured color/lightcurve objects are a subset of discovered objects.} \label{fig:neo_pops}
\end{figure}

We find 1.27E5 NEOs are discovered in the ten-year survey: 1.1E5 from the $d \ge 10 \, \mathrm{m}$  population and 1.4E4 (upsampled from 3,026) from the undersampled $d \le 10 \, \mathrm{m}$ population. This represents a massive increase to the known $\sim$3.7E4 known NEOs \citep[retrieved 2024 December 2 from the SBDB;][]{JPL_tools}. Even if all known NEOs are among LSST's discoveries, we expect $\sim$9E4 new discoveries from LSST -- six times more than CSS \citep{christensen2012catalina,Jedicke2016}, the largest modern NEO discovery project with 16,112 NEO discoveries as of November 4, 2024 listed in the JPL Center for Near-Earth Object Studies database\footnote{\href{https://cneos.jpl.nasa.gov/stats/site_all.html}{https://cneos.jpl.nasa.gov/stats/site\_all.html}}. Our estimate is consistent with the order-of-magnitude prediction of $\sim$100,000 NEOs from the 2009 LSST Science Book \citep{LSST_book} and the upper-end of the the model-dependent estimate of 49,000-93,000 presented in \cite{dmtn-109}.

LSST's simulated discovery completeness on NEOs of different sizes is shown in Figure \ref{fig:neo_pops}. Cumulative completeness reaches 91\% for NEOs with $d >$1$\, \mathrm{km}$, a standard benchmark for the brightest and most observable objects. Among Potentially Hazardous Asteroids (PHAs), defined as having $d \ge 140 \, \mathrm{m}$ and minimum orbital intersection distances (MOID) under 0.05~au, LSST's completeness drops to 72.7\%, discovering 3152 of 4333 PHAs in our model. This is effectively the same as its MOID-independent 72.4\% completeness on $d >140\, \mathrm{m} $ NEOs. Differential completeness drops quickly at smaller sizes, though it does not reach zero within our model's diameter range. Still, the trend is clear -- only a handful of objects are expected to be found at meter and sub-meter scales. This result is broadly consistent with previous studies, in spite of them using different NEO population models, LSST cadences, image depths, and linking criteria: 75\% \citep{2007IAUS..236..353I}, 66\% \citep{ivezic2019lsst}, 62\% \citep{2016AJ....151..172G}, 66\% \citep{JONES2018181}, and 58-59\% \citep{verevs2017high}. There are two reasons for this consensus. One, the size distribution of PHAs is relatively well established (most remaining disagreement is at the low-size end of the NEO distribution). Second (and more important), this points to the robustness of Rubin's NEO discovery effectiveness to changes in observing cadence. Simply put, no matter the details of the cadence, as long as the entire visible sky is covered every 3-4 days, and there are two observations per night to construct tracklets, Rubin will end up discovering roughly 65-75\% of PHAs.

This is significant, as it re-confirms that LSST will make a substantial contribution towards the 2005 Congressional goal of 90\% PHA completeness \citep{USC_16691}. Only with contributions from existing major surveys and especially upcoming dedicated space missions \citep[NEO Surveyor;][]{mainzer2023near}, is it likely the community will reach the 90\% threshold by the late the 2030s.

\subsubsection{Arcs and Orbits}

\begin{figure*}
\begin{center}
\includegraphics[width=0.98 \textwidth]{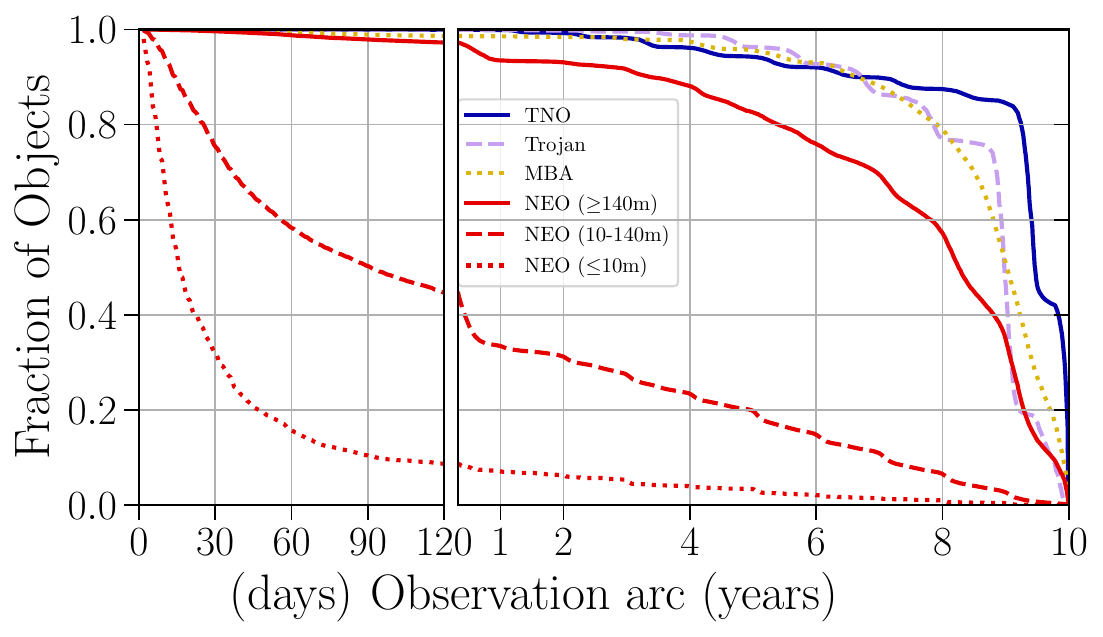}

\includegraphics[width=\textwidth]{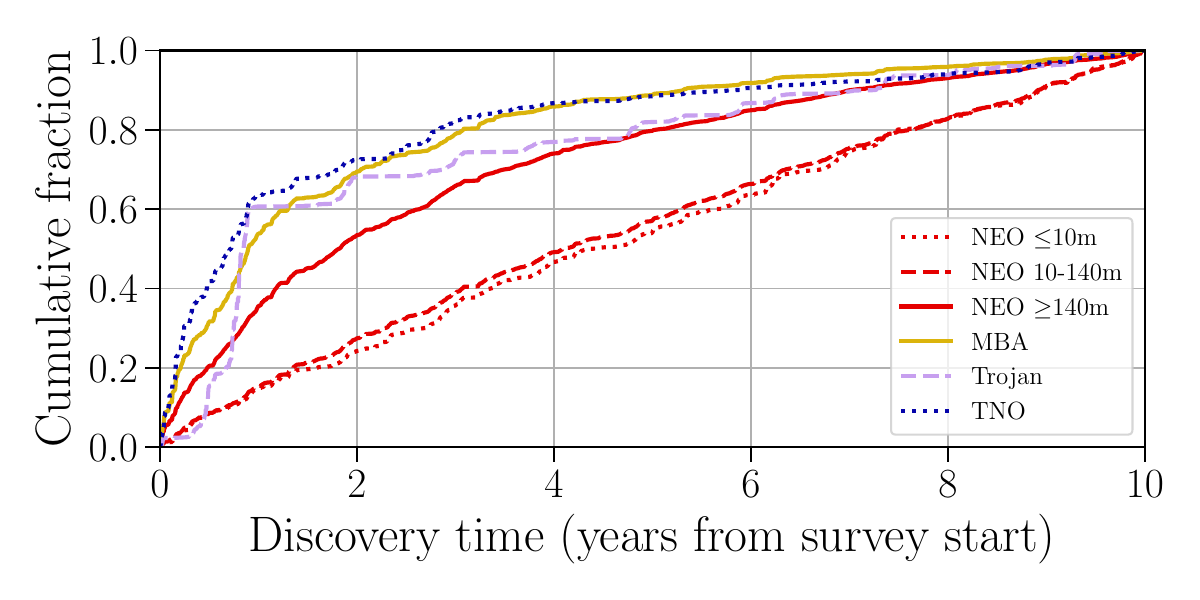}
\caption{The distribution of discovered input populations as a function of observation arc (top) and discovery time (bottom). Populations that are more distant and therefore move less over the ten-year survey, have longer observation arcs and their members tend to be discovered early. Conversely, small (and therefore nearby and fast-moving when detected) NEOs tend to be seen only for only one apparition and behave more like a transient population, though there is a sizable sample of $d \le 10 \, \mathrm{m}$ NEOs with long arcs. All slow-moving populations have highly concentrated observation arcs, a result of LSST's cadence. The Jupiter Trojan population, which is concentrated on the sky into an L4 and L5 cloud, has highly concentrated discovery times including a particularly large discovery spike in the first year of the survey when the larger L4 cloud is first observed. This makes the Trojans' observation arc distribution dependent on LSST's survey start date, unlike uniformly-distributed populations.}
\label{fig:dates}

\end{center}
\end{figure*}
We next look at observation arcs. These are typically a useful proxy for orbit fit quality \citep{doolittle1909simple, bernstein2000orbit,bowell2002asteroid}. For the purposes of this analysis, we define  objects with arcs over two years as being likely to achieve high-quality orbit fits. In total, 44,735 (35\%) discovered NEOs have observation arcs of more than two years -- a substantial improvement over the current situation where only 26\% of NEOs have arcs longer than two years \citep[retrieved 2025 Apr 7 from the SBDB;][]{JPL_tools}. For our NEOs, observation arcs are highly dependent on their size. As shown in Figure~\ref{fig:dates}, large NEOs have a fairly complete catalog of high-quality orbits: 13,023 $d>140\, \mathrm{m}$ objects (93\% of the discovered population, 67\% of the input population) are observed over at least two years, making it possible to place very tight constraints on their orbital distribution and terrestrial impact flux. Meanwhile, 90\% of $d<10\, \mathrm{m} $ objects have arcs less than three months, making rapid follow-up from the NEO community extremely time-sensitive to constrain small objects' orbits, colors, and lightcurves. The sample of 885 (6\%, upsampled from 200) $d<10\, \mathrm{m} $ NEOs with arcs of 2+ years will be suitable for very precise population studies of small NEOs, especially given Rubin's control of the systematics and the knowledge of LSST's selection function. Such a large population of small NEOs with high-quality orbits will help close the ``decameter gap" -- an order-of-magnitude discrepancy between the flux of $d<10\, \mathrm{m}$ impactors between telescopic and fireball observations \cite{ChowDecameter}. The LSST NEO sample will provide strong constraints on the NEO orbital distribution across multiple size regimes. 

There is also particular interest in investigating NEO disruption mechanisms at very low perihelion distances, which are evident through a steep deficit of NEOs on such orbits compared to dynamical models without disruption effects \citep{granvik2016}. Our catalog includes a 133 objects have $q < 0.2$ and at least 2 years of observation arc, compared to 79 currently known, and 1,619 with $q < 0.4$ compared to 579 currently-known \citep[retrieved 2024 December 5 from the SBDB;][]{JPL_tools}. This represents a threefold increase over the modern catalog of very low perihelion objects and a larger increase for moderately low perihelion objects, allowing more precise conclusions to be drawn once the survey is carefully debiased.

\subsubsection{Light Curves and Colors}

\begin{figure}
\begin{center}
\includegraphics[width=0.49 \textwidth]{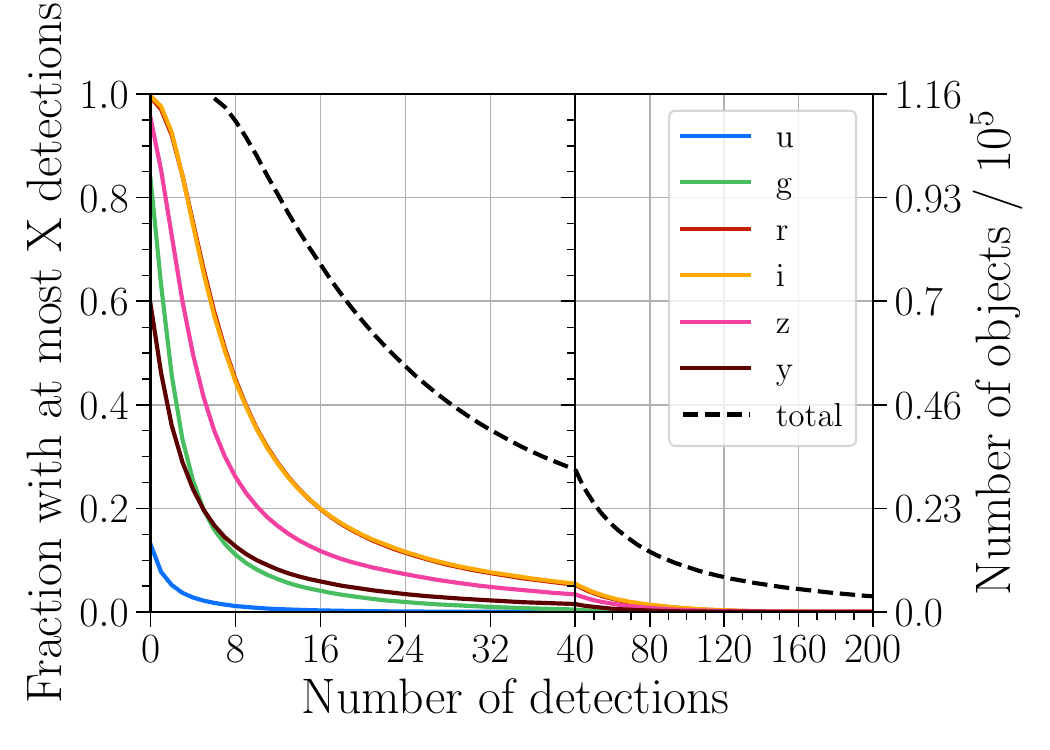}
\caption{Cumulative distribution of detections by band for NEOs. Note the change in x axis scale to display the tail of the overall detection distribution. Most objects receive only a few dozen detections, and small objects are not always detected in the \textit{u}, \textit{g}, or \textit{y} bands.}
\label{fig:neo_dets}
\end{center}
\end{figure}

We next examine the simulation's capacity for physical characterization. We present the distribution of NEO detections broken down by photometric band in Figure \ref{fig:neo_dets}. Typical LSST NEOs will have between 10 and 100 total detections, with significant dependency on size and apparent magnitude. The median discovered NEO has 23 detections, including zero $u$-band detections and one $y$-band detection. On the other hand, the median large ($d \ge 140\, \mathrm{m}$) NEO has 106 detections, including some in all $ugrizy$ bands. On the small end of the population, the median $d \le 10\, \mathrm{m} $ object has only 11 detections\footnote{This again illustrates the need for rapid follow-up of these objects, if their properties are desired to be understood.} across $griz$.

Our metric for high-quality $griz$ colors is met by 4.4E3 NEOs (3.5\%), including 4 ($\pm 4$; upsampled from 1) with $d <10\, \mathrm{m}$. The high-quality lightcurve criteria are met by 5E2 NEOs, almost all of which have $d \ge 140 \, \mathrm{m}$. In comparison, the recent NEOROCKS catalog \citep{birlan2024} includes colors for 170 NEOs, while the Asteroid Lightcurve Database \citep[LCDB; last updated 2023 October 1;][]{warner2009} lists 56 NEOs with derived shape models. It is clear that LSST's photometric measurements will enable the derivation of physical properties for a far larger catalog of objects than modern counterparts.

\subsubsection{Noteworthy Objects}

We also note certain discovered objects with extreme characteristics. Some NEOs pass through DDFs and are observed many more times than is typical in the WFD or NES. Our simulations include 16 such objects (all with $d \ge 10 \, \mathrm{m}$) which are detected more than 1,000 times. One object has been observed 2,647 times, enabling far more detailed characterization than would otherwise be possible purely from the main WFD survey. Some very small objects are discovered as well, including 216 ($\pm 28$; upsampled from 49) objects with $d \le 2~m$, and 4 ($\pm 4$; upsampled from one) with 1.0~m diameters. This sample will help constrain the NEO size distribution far beyond \texttt{NEOMOD3}'s debiased limit of 10~m.

\subsection{Main Belt Asteroids} \label{section:MBAs}
\begin{figure}
\begin{center}
\includegraphics[width=0.49 \textwidth]{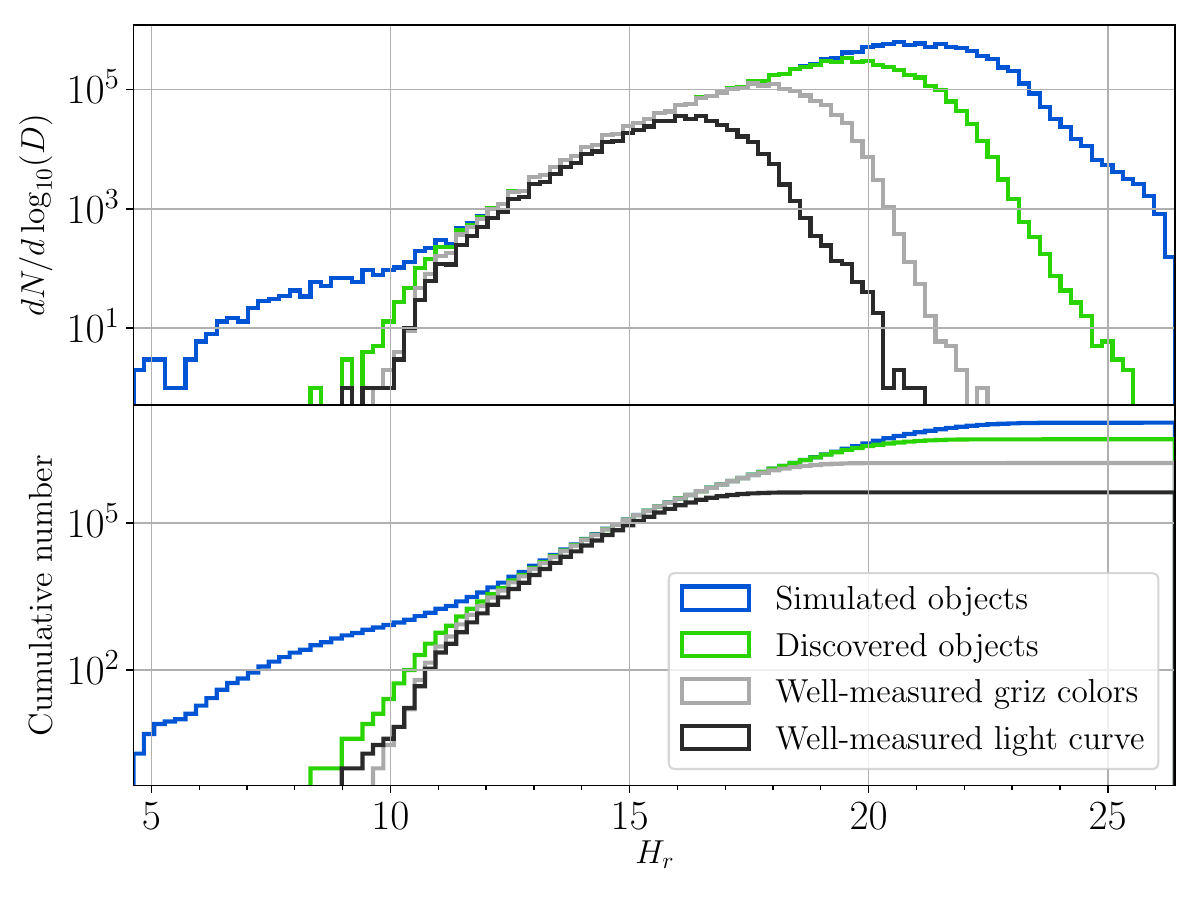}
\end{center}
\caption{The differential number density and cumulative number of MBAs (top and bottom panels respectively), analogous to Figure \ref{fig:neo_pops}. Loss of completeness at the bright end is due to the $m_r \sim 16$ saturation limit assumed for the LSST camera. Nearly all MBAs with $12.5 \lesssim H_r \lesssim 17$ will have well-determined colors, while about 75\% of objects with $12.5 \lesssim H_r \lesssim 15$ will have sufficient observations for high quality light curves.}
\label{fig:mba_pops}
\end{figure}

\subsubsection{Discovery Yield and Completeness} \label{mba:yield}

Our simulated LSST discovers 5.09E6 MBAs in the ten-year survey -- a factor of four more MBAs than the 1.34 million currently known \citep[retrieved 2024 December 2 from the SBDB;][]{JPL_tools}. Our number of MBA discoveries is very close to the LSST Science Book's estimate of 5.5 million and well within the 4.8-5.4 million predicted in \cite{dmtn-109}.

Our simulated survey reaches very high discovery completeness on sufficiently bright MBAs (see Figure \ref{fig:mba_pops}), all of which pass through the WFD and NES survey areas during the ten-year survey and exhibit sufficiently slow on-sky motions to be easily discoverable. Discovery completeness of 100\% is reached for Jupiter Trojans with $H_r$ between 8.7 and 14.9, and completeness is over 99\% for MBAs with $H_r$ between 12.2 and 18.5. We note that completeness drops to zero both on sufficiently large objects due to saturation (affecting only objects which have already been discovered by the many all-sky surveys to $m_r \sim 16$) and on sufficiently small ones.

\subsubsection{Arcs and Orbits} \label{mba:characteristics}

Observation arcs are very long for nearly all discovered MBAs, as shown in Figure \ref{fig:dates}. The median MBA arc is 9.0~yr, with 90\% longer than 6.6~yr, and 99\% longer than 1.0~yr. Taking arc as a proxy for quality of orbit fit, nearly all MBAs will have very precise orbital fits by the end of the survey, enabling exquisite understanding of asteroid families, their collisional structure, orbital perturbations, and possibly even allowing measurement of the Yarkovsky effect \citep[sunlight-induced non-gravitational acceleration;][]{Yarkovsky_LSST}.

\subsubsection{Light Curves and Colors} \label{section:mba_curves}

\begin{figure}

\begin{center}
\includegraphics[width=0.49 \textwidth]{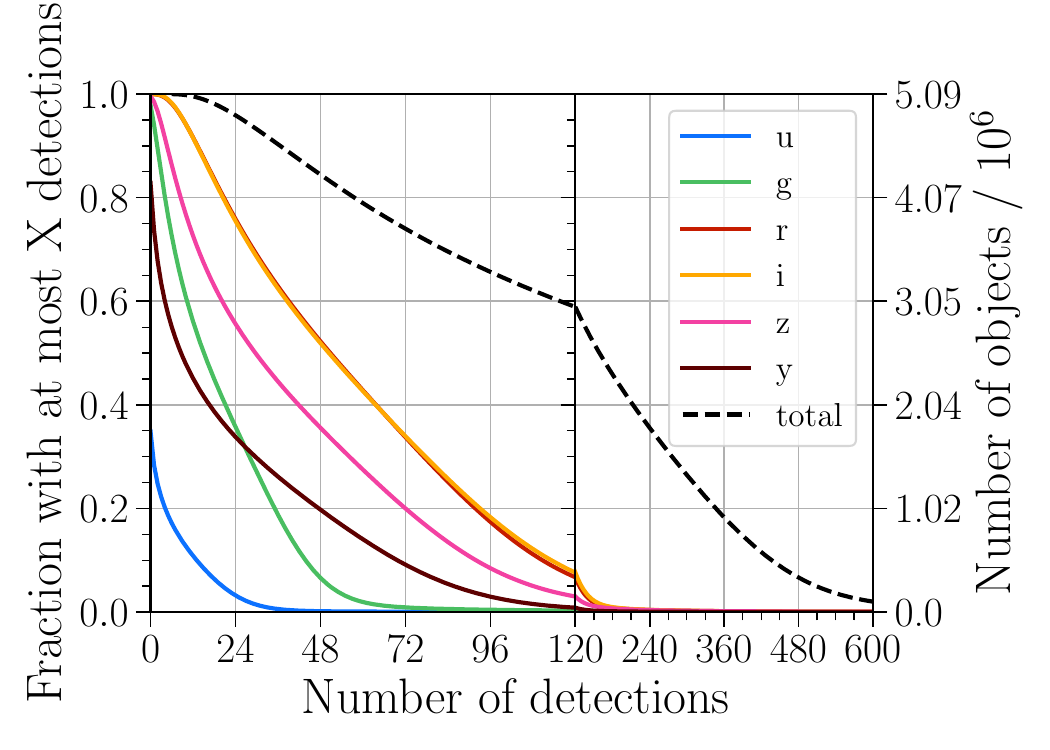}

\caption{Cumulative distribution of detections by band for MBAs, as in Figure~\ref{fig:neo_dets}. LSST will produce catalogs of Jupiter Trojans with hundreds of detections, and even the least-observed objects usually still have several detections in \textit{griz}.}\label{fig:mba_dets}
\end{center}

\end{figure}

We examine the photometric results of our simulations and apply our color and lightcurve metrics (see Section \ref{section:metrics}) to the catalog of MBA detections. The distribution of MBAs by numbers of detections per band is shown in Figure \ref{fig:mba_dets}. The median MBA is observed 160 times across all filters but $u$-. Predicting the populations of objects for which per-band lightcurves and shape models will be recoverable, we expect that 1.67E6 MBAs (32.8\%) will have high-quality light-curve-corrected \textit{griz} colors, over than 20 times more than the 72,043 objects with high-quality \textit{griz}  colors, mostly MBAs, provided by the Sloan Digital Sky Survey \citep{sergeyev2021million}, mostly MBAs. Meanwhile, light curve inversion will be possible for 4.21E5 MBAs (8.3\%), nearly 50 times more than the 8,600 MBAs included in \citet{durhan2023}. The massive color and lightcurve catalogs generated by a single well-characterized survey will enable characterization of the joint distribution of orbits, sizes, colors, and lightcurves in the style of \cite{bernardinelli2025photometry}, and enable the construction of a new generation of detailed debiased MBA population models based on a large-scale computation of accurate proper orbital elements \citep[e.g.,][]{nes2024}.
This will include revision of asteroid families inventory, especially the class of compact and young clusters, as well as an analysis of the spin states of their members and their correlation with the semimajor axis distribution predicted by the Yarkovsky evolution models \citep[e.g.,][]{vetal2015,durhan2023}.

\subsubsection{Noteworthy Objects}
Inner Solar System objects that pass through DDFs can be observed substantially more often than the main survey area's $\sim$800 coverings: 16,724 MBAs are detected at least 1,000 times each, with a single MBA reaching 4,923 detections. Uniquely precise characterization will be possible for such objects, such as the identification of super-fast rotation using the DDFs' rapid cadence \citep{Strauss_2024}. LSST will also enable the modeling of these populations to far smaller sizes than previously possible. There are 68 discovered MBAs with $H_V > 23.59$, the smallest (faintest) MBA currently known \citep[retrieved 2024 December 2 from the SBDB;][]{JPL_tools}, including a smallest discovered object with $H_V = 24.07$.
\subsection{Jupiter Trojans}\label{section:Trojans}
\begin{figure}
\begin{center}
\includegraphics[width=0.49 \textwidth]{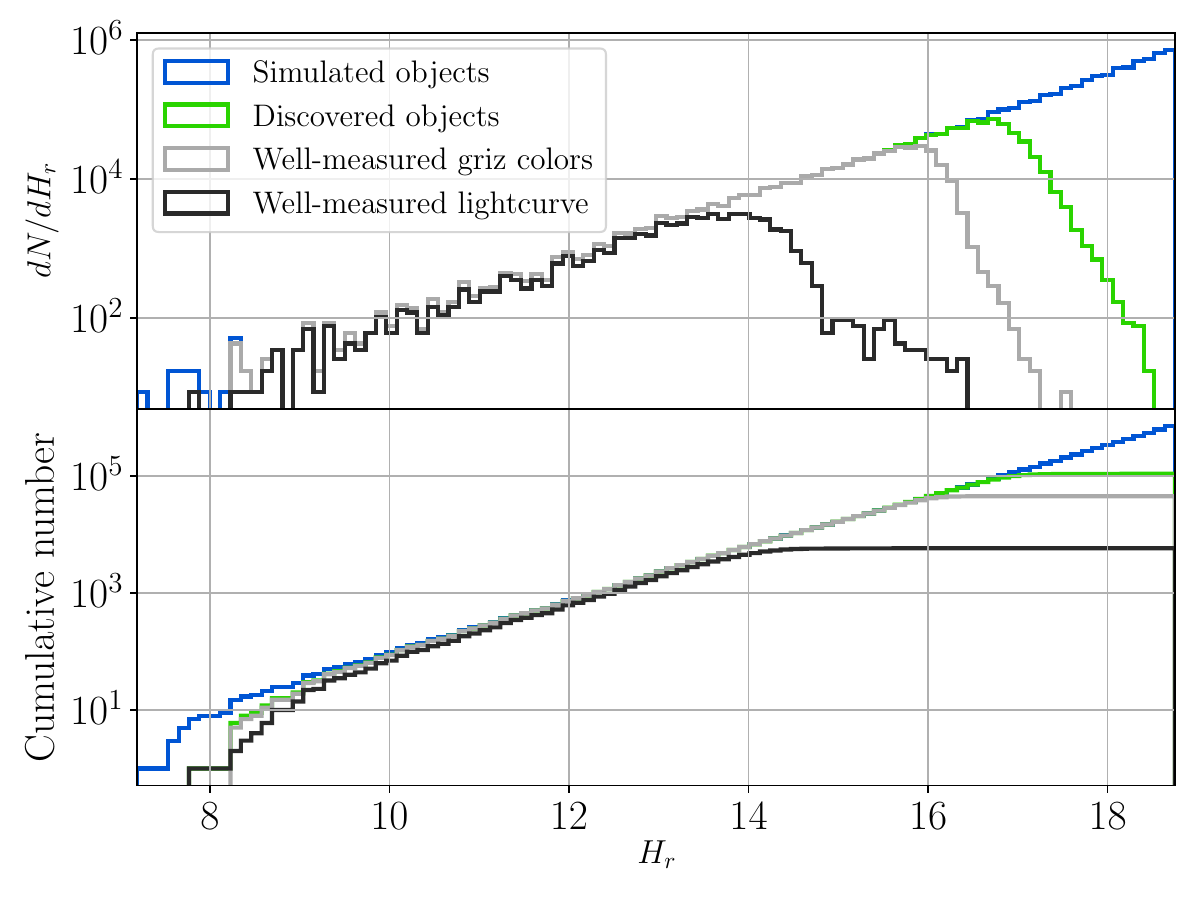}
\end{center}
\caption{The differential number density and cumulative number of Jupiter Trojans (top and bottom panels respectively). The Trojan population is generally very well-measured, with large size ranges receiving high discovery completeness, color completeness, and lightcurve completeness.}
\label{fig:trojan_pops}
\end{figure}
\subsubsection{Discovery Yield and Completeness}
Our simulated LSST discovers 1.09E5 Jupiter Trojans in the ten-year survey -- a factor of seven more Jupiter Trojans than the 15,134 currently known \citep[retrieved 2025 April 1 from the SBDB;][]{JPL_tools}. Our number of Jupiter Trojan discoveries is substantially lower than the prediction of 280,000 \citep{LSST_book}, primarily a result of modern updates to Jupiter Trojan population models. 

Our simulated survey reaches very high discovery completeness on sufficiently bright Jupiter Trojans (see Figure \ref{fig:trojan_pops}), which pass through the WFD and NES survey areas during the ten-year survey and exhibit sufficiently slow on-sky motions to be easily discoverable. Discovery completeness of 100\% is reached for Jupiter Trojans with $H_r$ between 8.5 and 15.5, and discovery efficiency converges to zero for bright and dim objects due to saturation and the limiting magnitude respectively.

\subsubsection{Arcs and Orbits}
Observation arcs are very long for nearly all discovered Jupiter Trojans, as shown in Figure \ref{fig:dates}. The median 9.0~yr, with 90\% longer than 6.7~yr, and 99\% longer than 3.6~yr. Taking arc as a proxy for quality of orbit fit, nearly all Jupiter Trojans will have very precise orbital fits by the end of the survey.

\subsubsection{Light Curves and Colors}
\begin{figure}
\begin{center}
\includegraphics[width=0.49 \textwidth]{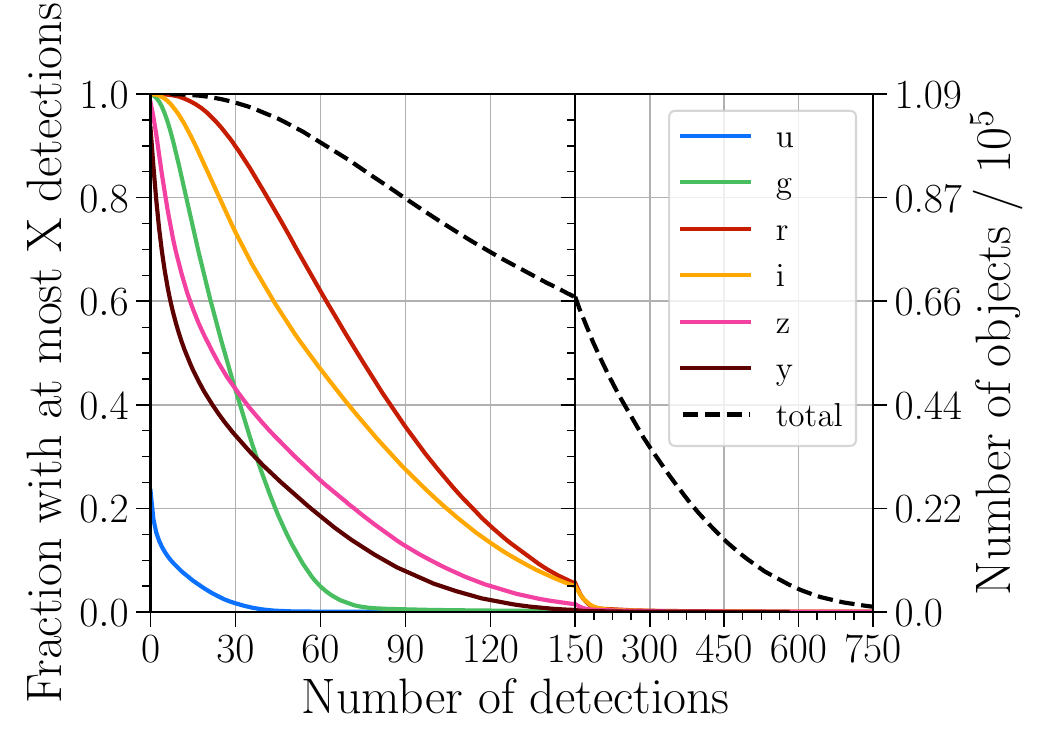}

\caption{Cumulative distribution of detections by band for Trojans, as in Figure~\ref{fig:neo_dets}. LSST will produce catalogs of Jupiter Trojans with hundreds of detections (median , and even the least-observed objects usually still have several detections in \textit{gri}.}\label{fig:tr_dets}
\end{center}
\end{figure}
We examine the photometric results of our simulations and apply our color and lightcurve metrics (see Section \ref{section:metrics}) to the catalog of Jupiter Trojan detections. The distribution of Jupiter Trojans' numbers of detections per band is shown in Figure \ref{fig:tr_dets}, with the median Trojan detected 197 times including observations in each band except $u$-. Predicting the populations of objects for which per-band lightcurves and shape models will be recoverable, we expect that 4.5E4 Jupiter Trojans (41.3\%) will have high-quality light-curve-corrected \textit{griz} colors, a massive expansion over the 186 Trojans similarly measured by DES \citep{DES_Trojans}. Light curve inversion will be possible for 5.8E3 Jupiter Trojans (5.3\%), dwarfing the 79 Jupiter Trojans with shape models reported by \citet{Hanus2023}. The large color and lightcurve catalogs generated by a single well-characterized survey will enable characterization of the joint distribution of orbits, sizes, colors, and lightcurves and joint orbit-size-color-lightcurve populations models.

\subsubsection{Noteworthy Objects}
Jupiter Trojans passing through DDFs can be observed substantially more than the main survey area's $\sim$800 coverings, including 321 Jupiter Trojans detected at least 1,000 times each and a single Jupiter Trojan reaching 4,664 detections. As with well-observed MBAs, precise characterization will be possible for these objects, such as the identification of super-fast rotation using the DDFs' rapid cadence \citep{Strauss_2024}. LSST will also probe the Trojan populations to far smaller sizes than previously possible: our catalog includes 28 discovered Trojans with $H_V \ge 18.38$, that of the smallest (faintest) Trojan currently known \citep[retrieved 1 April 2025 from the SBDB;][]{JPL_tools}, and a smallest discovered object with $H_V = 18.74$. The hundreds of discovered Jupiter Trojans with $H > 18$ will allow models to be calibrated nearly three magnitudes deeper than current models.

\subsection{The Outer Solar System (TNOs)}\label{section:TNOs}

\subsubsection{Discovery Yield and Completeness}

\begin{figure}

\begin{center}
\includegraphics[width=0.49 \textwidth]{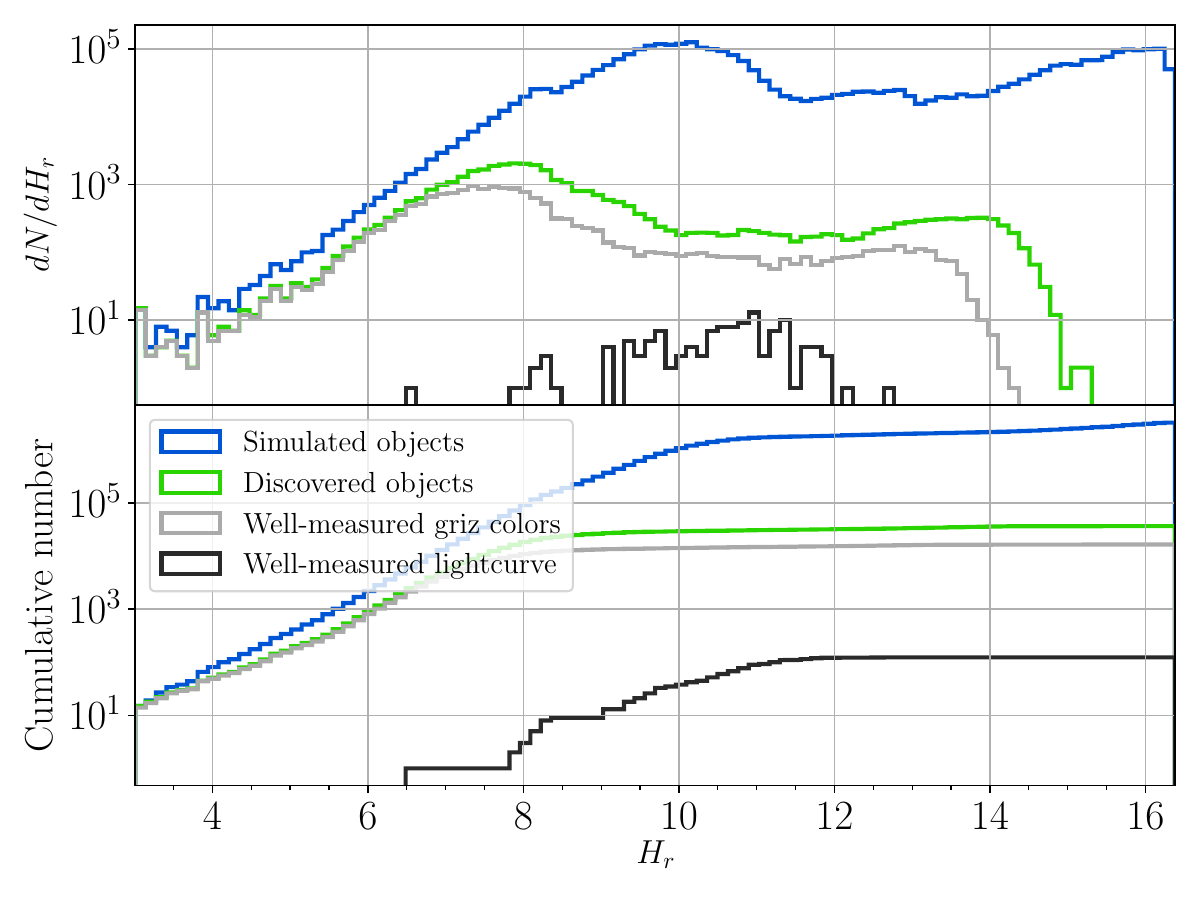}

\end{center}
\caption{Model TNOs by absolute magnitude, similar to Figure \ref{fig:neo_pops}. Due to the cut made on apparent magnitude at perihelion, the ``simulated objects'' population shown in blue does not strictly increase with $H_r$, despite our model having a strictly increasing size distribution.}\label{fig:tno_pops} 


\end{figure}

We find that 3.70E4 TNOs are discovered in the ten-year survey: 9,775 hot and 3,591 cold Classicals, 13,486 scattered disk objects, 2,543 detached objects, and 5,220, 207, 1,070, 448, and 662 objects (7,607 total) in the 3:2, 7:4, 2:1, 7:3, and 5:2 resonances, respectively. Our computed TNO yield is quite close to the 40,000 objects estimated in \cite{LSST_book}, and represents nearly an order of magnitude more objects than the currently-known 5,145 (as of 2024 December 2). More importantly, nearly all of these objects will be well characterized, both individually and as populations.  At the moment, the largest TNO samples with well-characterized selection function and physical properties come from OSSOS \citep[838 objects;][]{OSSOS_VII} and DES \citep[812 objects;][]{DES}; the LSST sample will deliver an increase of more than 40x over those datasets. Such a large well-characterized catalog will enable substantially more precise constraints on TNO populations than previous studies. The discovered and well-measured TNO samples' distributions by size are shown in Figure \ref{fig:tno_pops}.

The large sample of scattered disk objects (SDOs) that the LSST will discover is especially interesting as it will also include a large number of small, low-perihelion SDOs found much closer to the Sun than Neptune. The number of this sub-population is quite dependent on two poorly-constrained features of our model: the inner tail of the scattered disk's perihelion distribution and its size distribution down to $H_r \sim 15.2$ -- a small change in either would dramatically affect the population's size. Looking specifically at $H>9$ objects, whose discovery and orbital classification is listed as a high priority in \cite{roadmap}, our simulation indicates the LSST will discover 1,888 non-scattering and 8,585 scattering objects (largely with low perihelion) -- sample sizes sufficiently large to enable insights about the smallest-known TNOs, including some in all dynamical classes except cold classicals. However, while we do not model poorly-characterized populations like Sednoids or retrograde TNOs, the overall high completeness and discovery of some very distant objects (see Section \ref{section:noteworthy_tnos}) implies LSST will substantially expand the samples of these populations as well.

\begin{deluxetable}{c|ccccc}[h!]
\tabletypesize{\footnotesize}
	\tablecaption{$H_r$ completeness thresholds for each TNO subpopulation\label{table:tno_completeness}}
 \tablehead{\colhead{Population} & \colhead{Discovered} & \colhead{90\%} & \colhead{50\%} & \colhead{10\%} & \colhead{Limiting Completeness}}
 \startdata
 Cold Classicals & 3,591 & 6.61 & 7.51 & 8.01 & 100\% \\
 Hot Classicals & 9,775 & 6.71 & 7.7 & 8.45 & 84\% \\
 Detached & 2,543 & 2.07 & 5.46 & 7.73 & 76\% \\
 Scattering & 13,486 & 1.11 & 4.71 & 8.38 & 37\% \\
 Resonant (3:2) & 5,220 & 6.79 & 7.99 & 8.88 & 86\% \\
 Resonant (7:4) & 207 & 6.35 & 7.52 & 8.28 & 100\% \\
 Resonant (2:1) & 1,070 & 5.5 & 7.0 & 8.19 & 97\% \\
 Resonant (7:3) & 448 & 5.08 & 6.61 & 7.93 & 95\% \\
 Resonant (5:2) & 662 & 4.52 & 6.22 & 7.61 & 90\% 
\enddata
\end{deluxetable}

\begin{figure}

\begin{center}
\includegraphics[width=0.49 \textwidth]{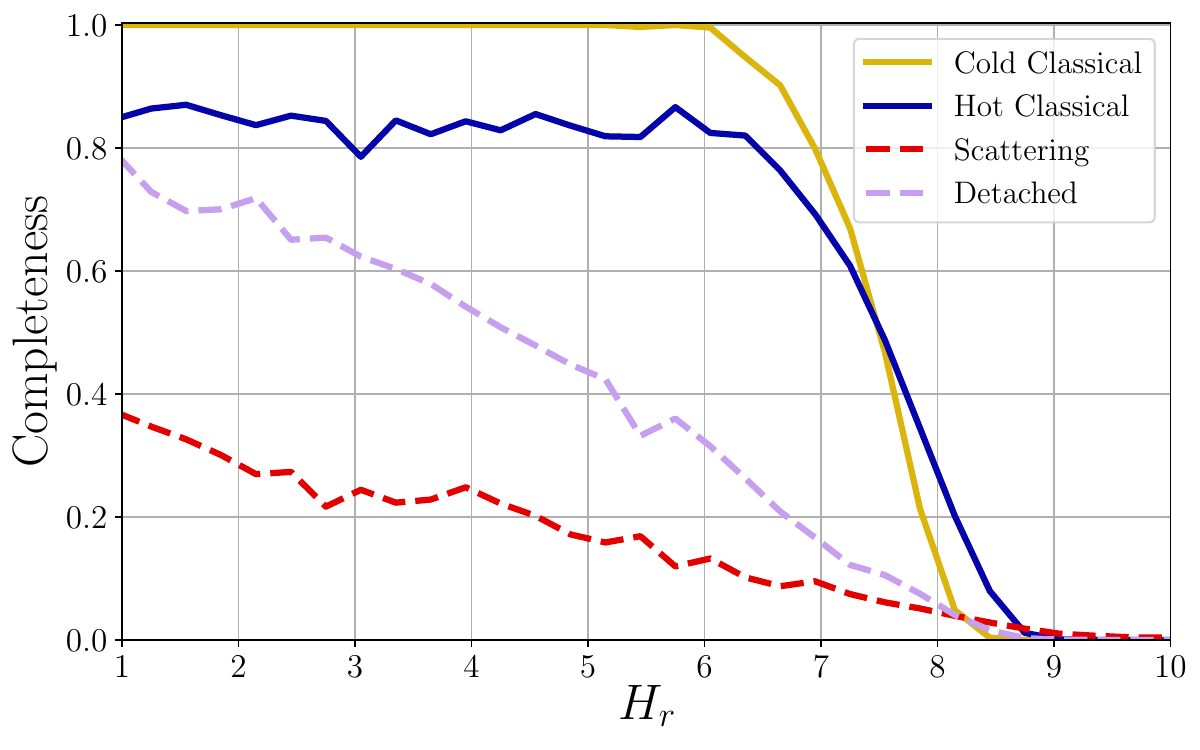} 

\end{center}
\caption{TNO discovery completeness by absolute magnitude for the non-resonant subpopulations. There is no size range on which most subpopulations have discovery completeness near 100\%. TNOs do not complete a large fraction of their orbits in ten years, so many eccentric scattering and detached objects which would be easily observed at perihelion stay at high distances throughout the survey. Meanwhile, high-inclination members of the hot classical population can stay further north than LSST observes for the entire survey. Cold classicals, with circular, low-inclination orbits, all lie within the survey footprint and reach high completeness for $H_r < 6$.}\label{fig:tno_completeness}


\end{figure}

Each population's limiting completeness on large objects is listed in Table \ref{table:tno_completeness} along with the $H_r$ at which completeness reaches 90\%, 50\%, and 10\% of that maximum. Completeness of the non-resonant populations are also shown in Figure \ref{fig:tno_completeness}. The resonant and classical populations, which have fairly constrained aphelia and inclinations, reach very high completeness up to $H_r \sim 5-7$. The scattering and detached populations, however, have aphelia too distant to be reliably detected even up to planetary sizes ($H_r \sim 0$), where objects near perihelion start to saturate LSST pixels and are no longer detected. This particular point differs greatly from estimates in \cite{LSST_book}, which argued that LSST would reach high completeness on sufficiently large scattered disk objects. We believe this is due to a different aphelion distribution used by that study. Nonetheless, even with low population completeness, discovering huge numbers of objects close to perihelion samples very effectively the detached and scattered populations.

\subsubsection{Arcs and Orbits}

Accurate determination of a TNO's orbit is required to understand its dynamical behavior, as the boundaries between the distinct TNO dynamical classes are quite narrow \citep{volk2024dynamical}, and typically multiple years of observing are required for high-quality dynamical classification. Given LSST's dense observing cadence, relatively uniform sky coverage, and the TNOs' slow on-sky motion, the majority of discovered TNOs will be observed throughout the survey. Specifically, we find that 99.8\% of TNOs will possess arcs greater than one year, 98\% will be greater than three years, with the median arc being a remarkable 9.5 years (see Figure \ref{fig:dates}). Therefore, by the end of LSST's nominal ten-year campaign, almost all discovered TNOs will have very well-constrained orbits and sufficient data for dynamical classifications. 

\subsubsection{Colors and Light Curves}
\begin{figure}
\begin{center}
\includegraphics[width=0.49 \textwidth]{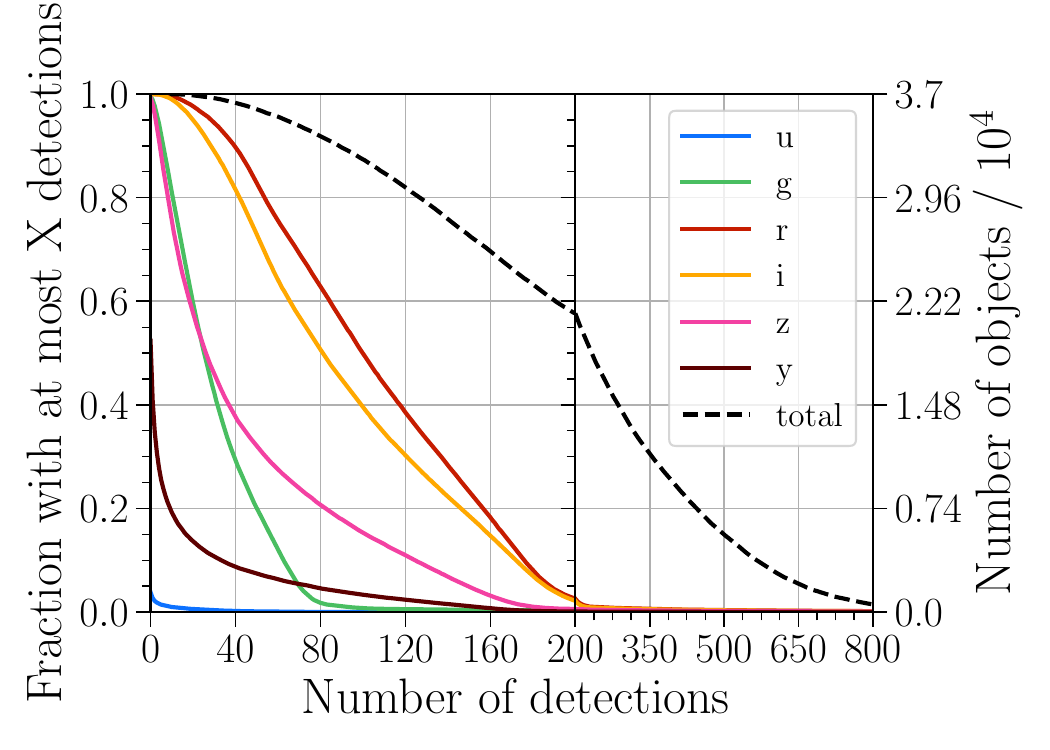}
\caption{Cumulative distribution of detections by band for TNOs, analogous to Figure~\ref{fig:neo_dets}.}\label{fig:tno_dets}
\end{center}

\end{figure}

We present distributions of TNO detections per band in Figure \ref{fig:tno_dets}. The median object receives 234 observations, mostly in the $r$- and $i$-bands, but with some contribution from $g$- and $z$-. Only the brightest TNOs receive $u$-band detections, and objects which lie in the area of the NES mini-survey, which images only in \textit{griz},  observations, have no $u$- and $y$-band detections regardless of brightness. As a result, fewer TNOs have $u$- and $y$-band detections than inner Solar System objects, which move faster and are guaranteed to leave the NES during the ten-year survey. 

Our metric for high-quality \textit{griz} colors (see Section \ref{section:metrics}) is met by 1.6E4 TNOs (45.0\%), a far larger population than any previous TNO color survey. For comparison, the Colours of the Outer Solar System Origins Survey (Col-OSSOS) measured the high-quality $g$, $r$, and $J$-band colors colors of only 102 objects \citep{Fraser_2023}. The TNO lightcurve metric is met by 1,057 TNOs (2.9\%), also a massive increase over the 148 TNOs with periods listed in LCDB \citep[last updated 2023 October 1;][]{warner2009} or the 19 with shape models from occultation \citep{TNO_occultations}. By generating these data products simultaneously with orbits, LSST will enable the investigation of TNOs' joint orbital-physical distribution and enable new insights into the history of the outer Solar System.

\subsubsection{Noteworthy Objects} \label{section:noteworthy_tnos}


A slow-moving TNO can spend far longer in a DDF than any faster-moving inner solar system object, resulting in far more measurements. These DDF-crossing TNOs include 218 with more than 1,000 detections, 18 with more than 5,000, and one object with 19,127 -- by far the most observations in our simulation. The high cadence of these observations, interleaved with the sparse imaging of the WFD, will enable especially accurate determination of objects phase and light curves, which can allow for the extraction of shape models, identification of contact binaries, finding pole orientations, and even constraining densities \citep{fernandez2022modeling}.

LSST's extensive coverage of high ecliptic latitudes will also enable the detection and characterization of objects with high inclinations ($i \gtrsim 40 \degr$). These rare objects are of particular interest for improving over current characterizations of the inclination distribution of the TNO population \citep{petit2017canada, bernardinelli2025photometry} and in particular differentiating between hypotheses of formation models of the outer Solar System \citep[\emph{e.g.}][]{Batygin_2016,Huang_2024,pfalzner2024}. Similarly, LSST will also enable a better characterization of the Sednoid population ($q \gtrsim 60 \, \mathrm{au}$). While we do not model this population directly, we find that some very distant TNOs are discovered by LSST, including six detached objects beyond $90$~au and one at $\sim 110$~au, indicating that we should expect some distant discoveries out towards the slow-moving limit of LSST's discovery pipelines (see Section \ref{section:discovery}). 

\section{Opportunities for Early Science} \label{section:early}
In the previous section we described the contents and opportunities provided by the full, 10-year, LSST catalog. However, a large fraction of this sample will become available after just a few years of survey time, opening opportunities for early solar system science.

We adopt LSST Data Release 3 (DR3) -- expected to be released around October 2028  \citep{guy2021rubin} -- as a convenient reference point for ``early science''. DR3 will include observations taken during the first two years of the survey\footnote{The first data release (DR1) will include data from the first 6 months, DR2 will include data from the first year, with subsequent releases continuing on an annual cadence through the final Data Release 11 (DR11)}, and be the first release with comprehensive all-sky templates and full Rubin solar system discovery capacity \citep{robinson2024tuninglegacysurveyspace}.

We find that most objects are discovered quite early in the ten-year survey, with distant populations quickly reaching high completeness (see Figure \ref{fig:dates}). By DR3, the LSST data set will already contain 72\% of TNOs, 68\% of Jupiter Trojans, 69\% of MBAs, and 53\% of $d >$140$\, \mathrm{m}$ NEOs expected to be discovered over the full 10-year survey. With the discovery of a large fraction of objects, accurate population estimates will be possible quickly, contingent only on survey characterization for debiasing. Also, research opportunities depending on new discoveries (i.e., detection of dynamically unusual objects, or identification of whole new populations) will be especially ripe in this early period. While we do not model Sednoids or other extreme TNO ($a > 250\, \mathrm{au}$, $q>37\,\mathrm{au}$) populations here, they will be discovered early with the other TNOs, allowing for an early re-evaluation of the evidence for the Planet 9 hypothesis \citep{Shankman2017,Bernardinelli2020,Napier2021,brown2021}. 

Unlike other populations, small NEOs are discovered more uniformly across the survey: only 27\% of NEOs with diameters $10$--$140 \, \mathrm{m}$ and 24\% of $d < 10\, \mathrm{m} $ NEOs will be found in the first two years. In the later years of the survey, the discovery rate settles to around $\sim 10,000$ per year, still more than three times the current discovery rate from CNEOS (as of 2024 December)\footnote{\href{https://cneos.jpl.nasa.gov/stats/site_all.html}{https://cneos.jpl.nasa.gov/stats/site\_all.html}}. Although LSST will provide a large number of NEO detections early in the survey, high-quality population characterization will be better conducted later in the survey. 

LSST's high precision color and light-curve metrics are enabled by its repeated, uniform covering of the entire sky. A quick estimation shows that in two years of WFD observations with the nominal filter distribution from \cite{jones2021survey}, objects must have SNR~$\sim$~60 in \textit{riz} (i.e. those already well-measured by other observatories) to meet our lightcurve metric, or SNR~$\sim$~13 in \textit{g} to meet our color metric. Higher-uncertainty colors from fewer detections will still be available early, but high-quality phase-corrected color measurements and shape-revealing light curves will only be available later in the survey.

\section{Summary and Conclusions} \label{section:conclusion}
Using the novel solar system survey simulation tool \texttt{Sorcha}, we generated a high-fidelity simulation of observations of NEO, MBA, Jupiter Trojan, and TNO populations for the LSST survey. We use this simulated catalog to predict the LSST discoveries yields for each population, and the numbers of high-quality orbits, colors, and lightcurves that will be delivered by LSST. We find that LSST will produce orbit, color, and lightcurve catalogs far larger than modern counterparts, enabling a new era of precision solar system science, including:

\begin{itemize}
    \item The LSST will independently link 5,356,423 objects (at least 3.9M of which will be new discoveries), including 1.27E5 NEOs, 5.09E6 MBAs, 1.09E5 Jupiter Trojans, and 3.70E4 TNOs -- 3-7 times more than are currently known in each category. We expect a total of 1.1 billion $5\sigma$ detections over the ten-year survey, more than twice as many as all previous surveys combined.
    \item Given this dataset, many presently uncertain features of small body populations will be quickly and strongly constrained, including the size distribution and number of Jupiter Trojans to $H_r < 18.5$, the size distribution of the scattering TNOs down to $H_r < 14$, and the sizes of extreme TNO populations.
    \item LSST's discovery completeness will be high for $d \ge $1$\, \mathrm{km}$ NEOs (91\%) and $d \ge $140$\mathrm{m}$ PHAs (72.7\%), confirming the finding of \cite{JONES2018181} that Rubin will significantly contribute to the global planetary defense effort. 
    \item Most objects detected in LSST will have long observational arcs, with median arcs of 9.0, 9.0, and 9.5 years for MBAs, Jupiter Trojans, and TNOs. The 44,735 (35\%) of NEOs with arcs longer than two years are significantly differentiated by size, but include 885 with $d \le 10 \, \mathrm{m}$.
    \item Median number of detections for NEOs, MBAs, Jupiter Trojans, and TNOs are 21, 160, 193, and 234, with $d >$140$\, \mathrm{m}$ NEOs receiving an order of magnitude more detections than $d <$10$\, \mathrm{m}$ ones. The $r-$ and $i-$ bands consistently contribute the most detections, while only a few objects receive $u-$ band detections. Most objects in all orbital classes are detected at least once in each $grizy$ band. 
    \item Objects which spend time in LSST's deep drilling fields (DDFs) receive detections at a far higher rate than those in the main survey area (the WFD). Bright, slow-moving objects will be detected there tens of thousands of times, enabling extremely precise light curve investigations including shape reconstructions.
    \item LSST's catalog of high-precision photometric data will be 10-300 times larger than present-day equivalents. We expect high-quality phase-corrected $griz$ colors for 4.4e3 (3.5\%) NEOs, 1.67E6 (32.8\%) MBAs, 4.5E4 Jupiter Trojans (41.3\%), and 1.6E4 (45.0\%) TNOs. Light curve inversions will be possible for 5E2 NEOs (0.03\%), 4.21E5 (8.3\%) MBAs, 5.8E3 Jupiter Trojans (5.3\%), and 1,057 (2.9\%) TNOs. 
    \item Much of this science will be possible {\em early} in the survey. As of DR3, expected to be released in late 2028, 72\% of LSST's TNOs, 68\% of its Jupiter Trojans, 69\% of its MBAs, and 53\% of its $d >$140$\, \mathrm{m}$ NEOs will be discovered. This sample will allow for population estimates nearly as precise as will be possible in the ten-year survey. High-quality colors and lightcurves will be available for some bright objects and those which pass through DDFs, but for most objects they will only become available later in the survey. 
\end{itemize}


The LSST has the potential to be transformational for our understanding of the small bodies of our Solar System, including their orbits, colors, shapes, and light curves. As these serve as tracers and records of planet formation and evolution, insights derived from Rubin's data will add to our understanding of how our and other planetary systems came to be.

Importantly, while the data collection for LSST will last for a decade, we have shown here that the majority of new object discoveries will come rapidly -- as quickly as just two years into the survey -- enabling major, important, science questions to be tackled early on. Given Rubin solar system discoveries and observations will be public by virtue of daily submissions to the Minor Planet Center, we hope these insights will encourage the broad community to get ready and use the LSST dataset.
\\

The analyses we've made here serve only to illustrate the potential of the LSST catalog; many more investigations are both possible and desirable. To support them, we have made the simulated detection catalogs public, available at \href{https://www.canfar.net/citation/landing?doi=25.0062}{https://www.canfar.net/citation/landing?doi=25.0062}. We also provide the input populations as well as instructions on how to rerun the simulation, if needed. We hope this will allow the readers to test whether their particular science cases could benefit from the LSST, and get ready for this extraordinary dataset.

\section{Acknowledgements}
We thank our anonymous reviewer for their clear, precise feedback, which improved the quality of this work.

J.A.K. acknowledges the support from the University of Washington College of Arts and Sciences Department of Astronomy and thanks the LSST-DA Data Science Fellowship Program, which is funded by LSST-DA, the Brinson Foundation, and the Moore Foundation; his participation in the program has benefited this work.

J.A.K., P.H.B., and M.J. acknowledge support from the DIRAC Institute in the Department of Astronomy at the University of Washington. The DIRAC Institute is supported through generous gifts from the Charles and Lisa Simonyi Fund for Arts and Sciences, and the Washington Research Foundation.

M.E.S acknowledges support through the LSST Discovery Alliance LINCC Frameworks Incubator grant [2023-SFF-LFI-01-Schwamb]. Support was provided by Schmidt Sciences. S.R.M. and M.E.S. acknowledge support in part from UK Science and Technology Facilities Council (STFC) grants ST/V000691/1 and ST/X001253/1. 

J. Murtagh acknowledges support from the Department for the Economy (DfE) Northern Ireland postgraduate studentship scheme and travel support from the STFC for UK participation in LSST through grant ST/S006206/1.

G.F. acknowledges support in part from STFC grant ST/P000304/1. This project has received funding from the European Union’s Horizon 2020 research and innovation program under the Marie Sk\l{}odowska-Curie grant agreement No. 101032479.

S.E. and S.C. acknowledge support from the National Science Foundation through the following awards: Collaborative Research: SWIFT-SAT: Minimizing Science Impact on LSST and Observatories Worldwide through Accurate Predictions of Satellite Position and Optical Brightness NSF Award Number: 2332736 and Collaborative Research: Rubin Rocks: Enabling near-Earth asteroid science with LSST NSF Award Number: 2307570.

D.V. acknowledges support from the Czech Science Foundation through grant 25-16507S.

This material or work is supported in part by the National Science Foundation through Cooperative Agreement AST-1258333 and Cooperative Support Agreement AST1836783 managed by the Association of Universities for Research in Astronomy (AURA), and the Department of Energy under Contract No. DE-AC02-76SF00515 with the SLAC National Accelerator Laboratory managed by Stanford University.  

This research has made use of NASA’s Astrophysics Data System Bibliographic Services. This research has made use of data and/or services provided by the International Astronomical Union's Minor Planet Center. The SPICE Resource files used in this work are described in \citet{acton1996, acton2018}. Simulations in this paper made use of the REBOUND N-body code \citep{rein2012}. The simulations were integrated using IAS15, a 15th order Gauss-Radau integrator \citep{rein2015ias15}. This work made use of Astropy:\footnote{http://www.astropy.org} a community-developed core\texttt{python} package and an ecosystem of tools and resources for astronomy \citep{astropy:2013,astropy:2018,astropy:2022}. 

This work made use of the following software packages: \texttt{Sorcha} \citep{Sorcha, Sorcha_Matt}, \texttt{astropy} \citep{astropy:2013, astropy:2018, astropy:2022}, \texttt{Jupyter} \citep{2007CSE.....9c..21P, kluyver2016jupyter}, \texttt{matplotlib} \citep{Hunter:2007}, \texttt{numpy} \citep{numpy}, \texttt{pandas} \citep{mckinney-proc-scipy-2010, pandas_6702671}, \texttt{python} \citep{python}, and \texttt{scipy} \citep{2020SciPy-NMeth, scipy_5725464}. Some of the results in this paper have been derived using \texttt{healpy} and the HEALPix package\footnote{http://healpix.sourceforge.net} \citep{Zonca2019, 2005ApJ...622..759G, healpy_5012376}.

We gratefully acknowledge the support of the Center for Advanced Computing and Modelling, University of Rijeka (Croatia), for providing supercomputing resources at HPC Bura. Computations in this paper were run on the FASRC Cannon cluster supported by the FAS Division of Science Research Computing Group at Harvard University. We thank the Canadian Advanced Network for Astronomical Research (CANFAR) for making our data publicly available.

\bibliography{refs.bib}
\end{document}